# Moiré Phonons and Emergent Exciton-Phonon Coupling in a Moiré Heterobilayer

Can B. Uzundal[1,2,3,4,*,†], Woochang Kim[1,4,*], Zhiyuan Cui[1,4*], Yuxuan Wei[1,4], Zheyu Lu[1,4,5], Qixin Feng[1,4], Francis L. Hong[1,#], Indrajit Maity[6,7], Takashi Taniguchi[8], Kenji Watanabe[9], Manish Jain[6], Mit H. Naik[10,11], Yoseob Yoon[1,4,12], Michael F. Crommie[1,3,4], Steven G. Louie[1,4,†], and Feng Wang[1,3,4,5,†]

[1]Department of Physics, University of California, Berkeley, CA, USA
[2]Department of Chemistry, University of California, Berkeley, CA, USA
[3]Kavli Energy NanoScience Institute, Berkeley, CA, USA
[4]Materials Sciences Division, Lawrence Berkeley National Laboratory, Berkeley, CA, USA
[5]Graduate Group in Applied Science and Technology, University of California, Berkeley, CA, USA
[6]Centre for Condensed Matter Theory, Department of Physics, Indian Institute of Science, Bangalore 560012, India
[7]Department of Chemistry, Newcastle University, Newcastle upon Tyne, NE1 7RU, UK.
[8]Research Center for Materials Nanoarchitectonics, National Institute for Materials Science, 1-1 Namiki, Tsukuba 305-0044, Japan.
[9]Research Center for Electronic and Optical Materials, National Institute for Materials Science,1 1 Namiki, Tsukuba 305-0044, Japan.
[10]Department of Physics and Center for Complex Quantum Systems, University of Texas at Austin, Austin, TX, USA
[11]Oden Institute for Computational Engineering and Sciences, University of Texas at Austin, Austin, TX, USA.
[12]Department of Mechanical and Industrial Engineering, Northeastern University, Boston, MA, USA

*equal contribution

#present address: Department of Materials Science and Engineering, University of Illinois Urbana-Champaign, Urbana, IL 61801, USA

†Corresponding authors:

can_uzundal@berkeley.edu, sglouie@berkeley.edu, fengwang76@berkeley.edu,

**Abstract**

Moiré superlattices have emerged as a new platform for engineering electronic and optical properties in van der Waals heterostructures, enabling control over correlated and excitonic phenomena. Yet the impact of moiré superlattices on exciton-phonon coupling remains largely unexplored. Here we demonstrate emergent, layer-selective coupling between moiré phonons and moiré excitons in angle-aligned $WS_2$/$WSe_2$ heterobilayers. Using a broadband terahertz phonon transducer, we coherently launch moiré phonons that resonantly perturb the excitonic states. We show that the exciton-phonon coupling is intrinsically modified by the moiré superlattice in a layer-selective manner. A driven oscillator model captures the dynamics, revealing three moiré phonon resonances with distinct coupling to the moiré excitons. First principles calculations show that many moiré phonon modes can arise with distinct strongly hybridized in-plane and out-of-plane vibrations in the moiré unit cells. The calculations further identify the three experimentally observed moiré phonons and their emergent characteristic coupling to the moiré excitons.

## Main

Light-matter interactions in semiconductors are governed by excitons whose energy, lifetime and coherence are in general strongly influenced by interactions with free carriers, defects and lattice vibrations. Among these scattering processes, exciton-phonon coupling plays an important role in determining optical linewidths and relaxation dynamics[1–3]. While extensively studied in bulk semiconductors[4–7], the exciton-phonon coupling in van der Waals (vdW) moiré heterostructures has been little explored.

In vdW heterostructures, the moiré superlattice is accompanied by a strong three-dimensional lattice reconstruction[8–11] which reshapes both the exciton transition[12–18] and phonon vibrations of the moiré heterostructure[19–21]. This reconstruction gives rise to the emergence of multiple moiré exciton resonances[13,22–26] and moiré phonon modes[27–31]. Because both the exciton wavefunctions and phonon eigenmodes vary spatially within the moiré unit cell, the moiré exciton-phonon coupling can be distinct from conventional semiconductor systems.

In this work, we directly probe moiré exciton-phonon coupling in angle-aligned $WS_2$/$WSe_2$ heterobilayers using ultrafast pump-probe spectroscopy. Our time-domain approach enables layer-resolved, phase-sensitive tracking of phonon-induced exciton dynamics in both layers. We observe multiple moiré phonon resonances and find that moiré exciton–phonon couplings are enhanced relative to large twist angle samples. The moiré exciton-phonon couplings vary strongly for different moiré phonon and moiré exciton combinations. Complementary first-principles calculations identify the relevant low-energy moiré phonon modes and reveal that the selective exciton-phonon coupling is controlled by the layer-resolved in-plane lattice reconstruction of these hybrid shear-breathing modes, which governs how they modulate the moiré electronic structure.

**Experiment scheme**

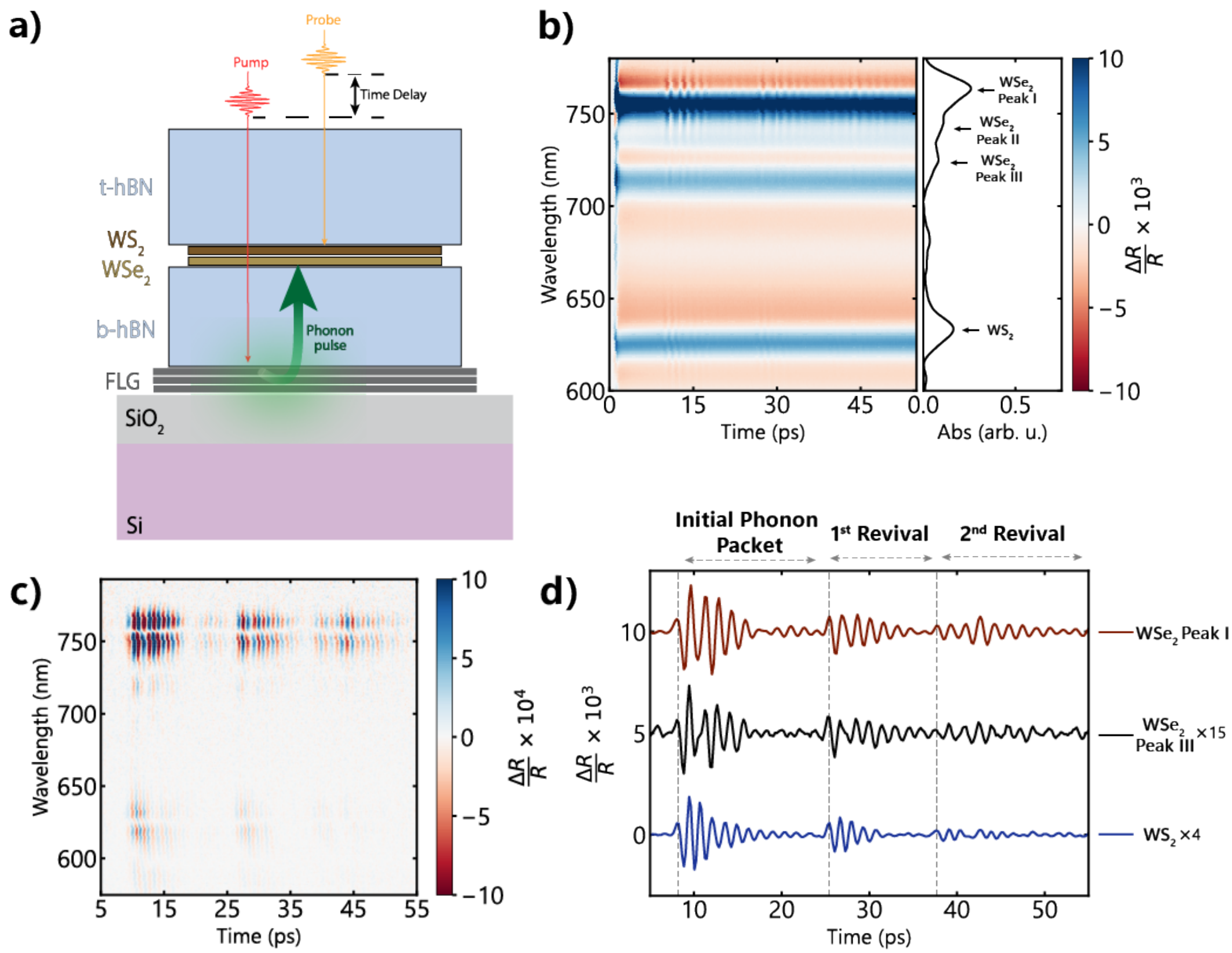

***Figure 1 a)*** *Experiment scheme. A short optical pump pulse induces expansion and contraction of few-layer graphene, generating a phonon wavepacket that propagates through the bottom hBN (b-hBN). The phonon wavepacket excites the moiré phonons in the $WS_2/WSe_2$ heterobilayer, which then modulate the moiré exciton reflectance through exciton-phonon coupling. A time-delayed supercontiuum probe pulse detects these reflectance changes.* ***b)*** *Right panel: Reflectivity of the $WS_2/WSe_2$ heterobilayer following a polynomial background subtraction, highlighting the relevant moiré excitons. Left panel: Raw pump-probe reflection contrast spectra. Strong coherent exciton-phonon coupling is observed for three moiré excitonic resonances in the angle-aligned $WS_2/WSe_2$ heterobilayer ($WSe_2$ Peak I, $WSe_2$ Peak III and $WS_2$).* ***c)*** *Transient reflection contrast spectrum after background removal. The signals are dominated by a coherent phonon-induced exciton shift, which are weak at each moiré exciton resonance, and changes sign between the regions above and below an exciton resonance.* ***d)*** *Isolated time-domain traces for each exciton region of interest. Traces are offset by $5x10^{-3}$ for clarity. The $WS_2$ trace is scaled by a factor of 4 and the $WSe_2$ Peak III trace by a factor of 15.*

Figure 1a illustrates our experimental scheme. We generate a coherent broadband terahertz phonon wavepacket using few-layer graphene (FLG) as a transducer. The broadband terahertz phonon wavepacket propagates through the hBN layers and excites the moiré phonons in the $WS_2/WSe_2$ heterostructure. The resulting modulation of the moiré excitons is measured using transient reflection spectroscopy with a time-delayed supercontinuum probe (550-800 nm) pulse.

Figure 1b shows a representative transient reflection measurement (left) together with the steady-state excitonic spectrum (right) of an angle-aligned $WS_2/WSe_2$ heterobilayer. We identify three moiré excitons associated with $WSe_2$ and one associated with $WS_2$. Following prior literature, we label these states $WSe_2$ Peak-I, II and III and the $WS_2$ moiré-exciton[22,25]. In contrast to the tightly bound $WSe_2$ Peak I-exciton, which is localized at the AA site of the moiré lattice, the $WSe_2$ Peak III-exciton arises from an intralayer charge-transfer exciton with the electron-hole pair delocalized across AA and AB regions[22,25]. These emergent excitonic states originate from three-dimensional lattice reconstruction and large strain distribution in the moiré superlattice.

The transient response in Fig. 1b exhibits three different dynamic components at each exciton wavelength. First, an ultrafast component exists only within the first few hundred femtoseconds around time zero due to the AC Stark effect[32,33]. Second, there is a slowly varying background extending over hundreds of picoseconds, which arises from pump-induced free carrier screening of the excitons. Finally, we observe a set of time-delayed coherent oscillations corresponding to moiré phonons excited by the phonon wavepacket in the heterobilayer.

To isolate the phonon-induced dynamics, we apply a high-pass Hann filter (passing frequencies above 0.3 THz) to the transient reflection data, resulting in the filtered spectra presented in Fig. 1c. The transient reflectivity modulation crosses zero near each moiré exciton feature and reverses sign on either side. Such a derivative-like line shape is indicative of a transient phonon-induced shift of the exciton resonance energy.

To extract the modulation amplitude from the moiré exciton-phonon coupling, we subtract the transient reflectivity modulation above and below each moiré exciton, improving the signal to noise ratio. The resulting coherent phonon dynamics for the $WSe_2$ Peak-I, WSe2 Peak-III and the $WS_2$ moiré-excitons are shown in Fig. 1d.

The data reveals several interesting behaviors. The coherent phonon-induced exciton modulation first appears ~8.5 ps after pump excitation, corresponding to the arrival of the phonon wavepacket from the few-layer graphene transducer. The signal persists for more than 10 ps and exhibits pronounced beating behavior. Another two revivals of the exciton modulation signal appear at 25.4 and 37.7 ps. These revivals originate from the arrival of phonon wavepackets reflected at the heterostructure interfaces. Using the revival timings together with the sound velocity in hBN (3.5 km/s), we extract the bottom and top hBN thickness of 29.8 nm and 51.6 nm, respectively[34–37].

To isolate the intrinsic exciton–phonon coupling, we focus on the initial phonon wavepacket and exclude the later arrivals associated with interface back reflections. We further restrict our discussion in the main text to the $WSe_2$ Peak I and the $WS_2$ moiré-exciton. The $WSe_2$ Peak II dynamics cannot be resolved due to overlapping spectral features and weak oscillator strength, while the $WSe_2$ Peak III exhibits qualitatively similar dynamics to $WSe_2$ Peak I-exciton. For completeness, the phonon-induced dynamics of the $WSe_2$ Peak III-exciton are discussed in the Supplementary Information ("Peak III Dynamics").

**Intrinsic Moiré Phonon-Induced Exciton Dynamics**

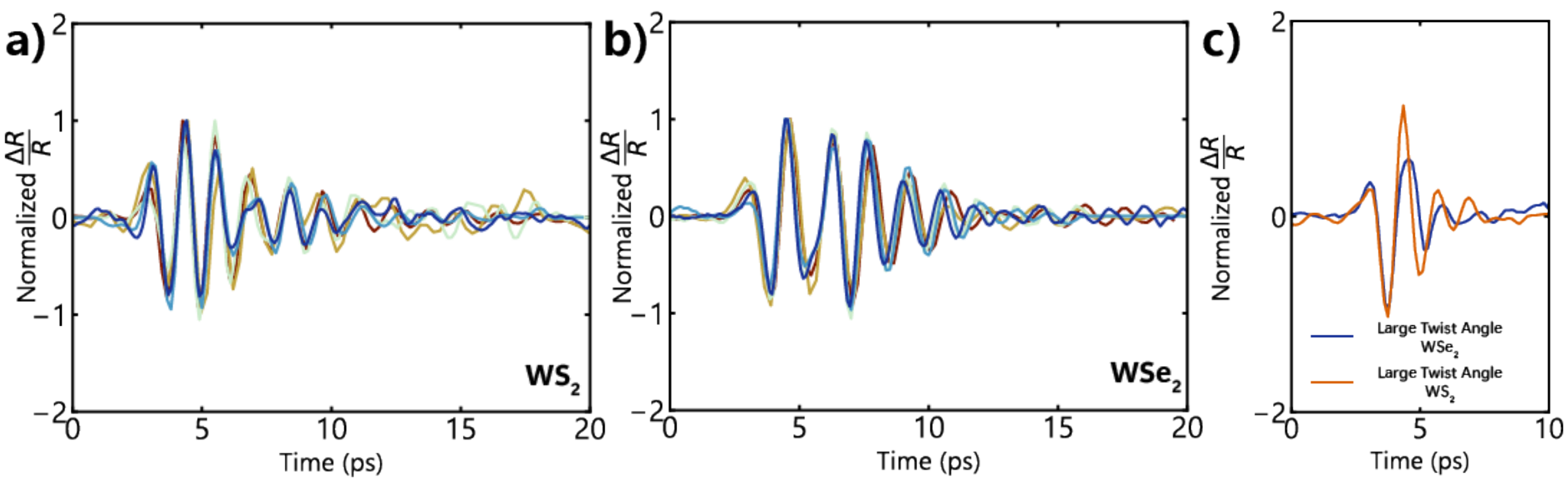

***Figure 2*** *Phonon-induced exciton dynamics observed in $WS_2/WSe_2$ heterobilayers.* ***a)*** *Time-domain phonon-induced exciton dynamics of the $WS_2$ exciton for four separate angle-aligned samples* ***b)*** *Corresponding $WSe_2$ Peak I-exciton dynamics.* ***c)*** *Phonon induced dynamics of $WS_2$ and $WSe_2$ excitons in a large twist angle heterobilayer. Each of the colored curves are normalized to have a peak value of 1 to emphasize the qualitative differences. For visual comparison, each time trace is shifted by a constant offset that accounts for differences in the bottom hBN thickness, with an additional small offset applied to highlight the onset of the phonon-induced response.*

Figure 2a and b show phonon-induced exciton dynamics measured across four angle-aligned $WS_2/WSe_2$ heterobilayers with different device parameters. Device details are summarized in the Supplementary Information ("Device parameters"), with microscope images presented in Fig. S2 and twist angles determined from second harmonic generation anisotropy in Fig. S3. A key requirement for obtaining reproducible dynamics is the use of thick top and bottom hBN layers (thickness > 30nm), which allows for clear separation of the first phonon wavepacket from the subsequent back reflected wavepackets. This reproducible behavior contrasts with prior measurements on the same moiré system[38], where the presence of multiple transducers and overlapping multiple reflections (i.e. Fabry–Pérot effects) obscured the intrinsic dynamics. (See Fig. S4 and the Supplementary Information ("Fabry–Pérot Effects") for a more detailed discussion).

To examine the role of the moiré superlattice, we performed control measurements in a large twist angle $WS_2/WSe_2$ heterobilayer with negligible moiré effects. The data is shown in Fig. 2c. Compared with the angle-aligned $WS_2/WSe_2$ heterostructure, the phonon-induced modulation in the $WS_2$ and $WSe_2$ excitons has a much shorter lifetime and a smaller amplitude (see Fig. S5 and Supplementary Information ("Time Domain Comparison") for a direct comparison). Similarly, the exciton-phonon coupling in natural bilayer $WSe_2$ is also weak and short lived (see Fig. S6 and Supplementary Information ("Natural Bilayer")). Our results establish that the moiré superlattice can lead to extended moiré phonon lifetime and enhanced moiré exciton-phonon couplings in angle-aligned moiré heterostructures.

**Spectra and Phase of the Moiré Phonon-Induced Exciton Modulation**

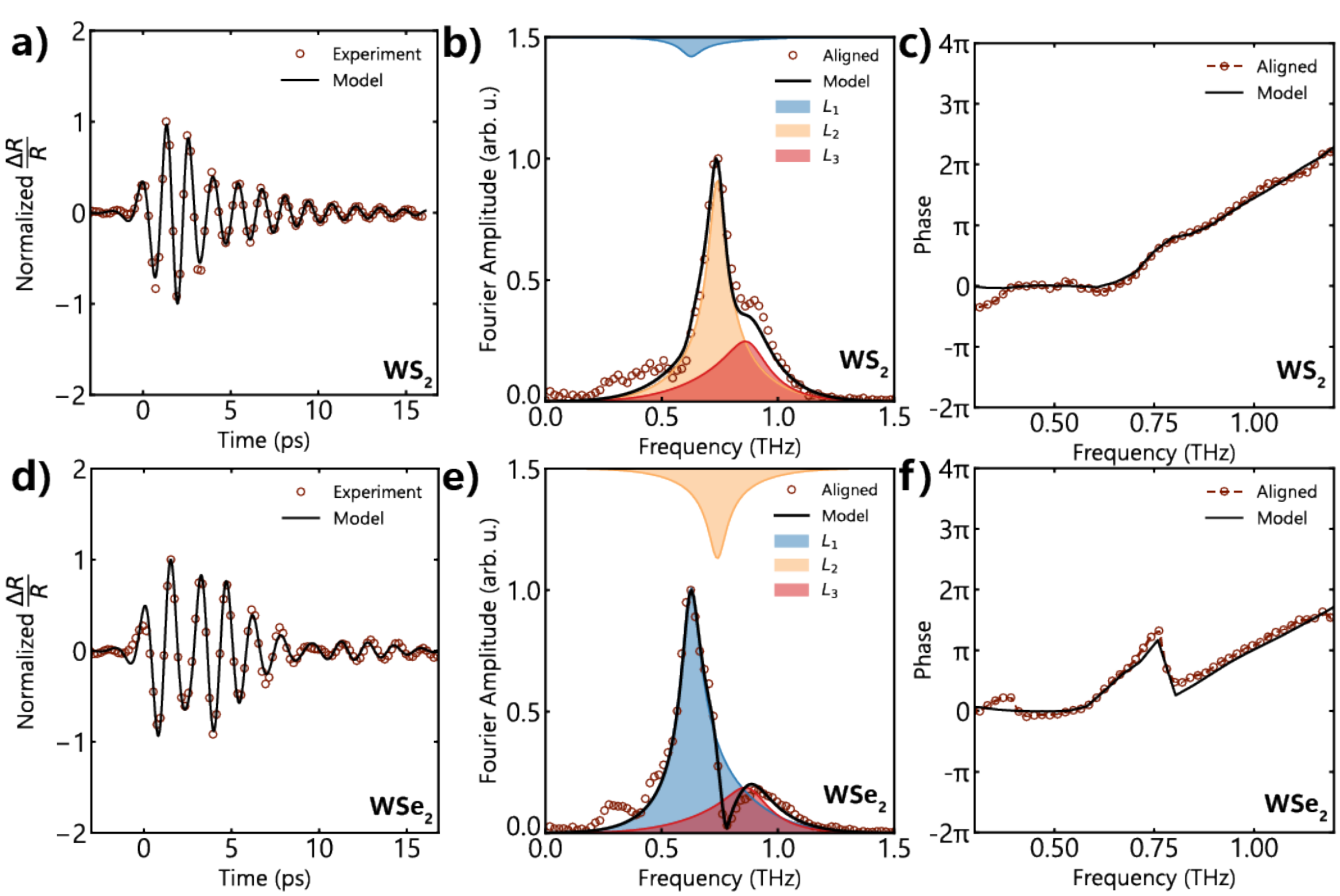

***Figure 3 a)*** *Phonon-induced exciton dynamics of the $WS_2$ A, the black line shows the three driven oscillator model fit.* ***b)*** *Corresponding Fourier amplitude spectrum. Colored regions indicate contributions of individual Lorentzian components,* $L_k^{(i)}(\omega)U(\omega)$, *where k denotes the* $k^{th}$ *oscillator and i is a layer index; negative oscillator strengths are shown as inverted peaks.* ***c)*** *Frequency dependent relative phase of the $WS_2$ phonon response.* ***d)*** *Phonon-induced exciton dynamics of the $WSe_2$ Peak I, the black line shows the model fit.* ***e)*** *Fourier amplitude spectrum. Colored regions indicate contributions of individual Lorentzian components; negative oscillator strengths are shown as inverted peaks.* ***f)*** *Frequency dependent relative phase of the $WSe_2$ phonon response.*

Time domain phonon spectroscopy is a phase sensitive measurement that can reveal hidden coupling mechanisms in van der Waals materials[39]. Here we analyze the layer-resolved coherent dynamics by Fourier transforming the time-domain traces for $WSe_2$ and $WS_2$ moiré excitons. Figure 3a presents the phonon-induced exciton dynamics of the $WS_2$ moiré exciton, which we denote as $H_{WS_2}(t)$. Its Fourier transform, $H_{WS_2}(\omega) = \int H_{WS_2}(t)e^{-i\omega t}dt$ yields the amplitude spectrum $|H_{WS_2}(\omega)|$ (red circles in Fig. 3b) and the phase $\theta(\omega) = \arg[H_{WS_2}(\omega)]$ (red circles in Fig. 3c). We present the phase after removing the trivial linear-in-frequency contribution associated with the choice of time zero.

Qualitatively, Fig. 3b shows that $WS_2$ moiré exciton couples to a dominant oscillation at 0.74 THz, with a weaker shoulder at slightly higher frequency. In contrast, the Fourier spectrum of the large-twist-angle heterobilayer lacks these sharp resonances (see Fig. S8 for a direct comparison). This contrast highlights the emergence of well-defined resonances in the aligned $WS_2/WSe_2$ heterobilayer.

Such resonances are also reflected in the phase response; a Lorentzian oscillator produces a characteristic $\pi$ phase shift as the frequency is swept across a resonance. Consistent with this behavior, Fig. 3c shows a pronounced phase shift near 0.74 THz, followed by a broader and more subtle feature at higher frequencies. By comparison, the large-twist-angle sample remains mostly featureless across the same spectral range.

We apply the same analysis to the $WSe_2$ Peak I-exciton of the aligned $WS_2/WSe_2$. Figures 3d–f present the corresponding time-domain trace, amplitude spectrum, and phase. The spectrum exhibits a dominant peak at 0.62 THz, a pronounced dip near 0.74 THz and a secondary shoulder at higher frequencies. Together with the $WS_2$ resonances, these frequencies fall within the expected range for layer breathing and shear modes in bilayer TMDs . The phase response (Fig. 3f) exhibits strong dispersive features associated with these resonances, including a phase jump near 0.74 THz, while large-twist-angle sample again shows no comparable structure.

**Driven Oscillators Model of Moiré Exciton-Phonon Coupling**

We model the dynamics using a multiple driven-oscillator framework that captures the time-domain response. The details of the model and the fitting routine are provided in the supplementary information ("Phonon Model"). Briefly, a phonon wavepacket with Gaussian envelope $U(\omega)$ drives the system through an effective susceptibility $\chi_{eff}^{(i)}(\omega)$ , where $i \in \{WS_2, WSe_2\}$. The complex frequency-domain response is given by

$$H_{model}^{(i)}(\omega) = \chi_{eff}^{(i)}(\omega)\, U(\omega)$$

and the corresponding time-domain signal is obtained via the inverse Fourier transform

$$H_{model}^{(i)}(t) = \mathcal{F}^{-1}\left[H_{model}^{(i)}(\omega)\right]$$

We directly compare $H_{model}^{(i)}(t)$ to the measured traces (Fig. 3a, d) to extract the parameters of $\chi_{eff}^{(i)}(\omega)$. The effective susceptibility is modeled as a sum of three Lorentzian oscillators,

$$\chi_{eff}^{(i)}(\omega) = \sum_{k=1}^{3} L_k^{(i)}(\omega), \quad L_k^{(i)}(\omega) = \frac{d_k^{(i)}\, \omega_k^2}{\omega_k^2 - \omega^2 - i\gamma_k\omega}$$

where $d_k^{(i)}$ is an effective oscillator strength, $\omega_k = 2\pi f_k$ the resonance frequency, and $\gamma_k$ the linewidth. We employ a joint fitting procedure in which the same $\omega_k$ and $\gamma_k$ are used to describe the moiré phonon resonances responsible for the transient signals in both $WS_2$ and $WSe_2$ layers, but the oscillator strengths $d_k^{(i)}$ are allowed to vary independently. The resulting fits (black curves in Fig. 3) reproduce the measured dynamics in both layers, including the detailed phase evolution across the resonances.

From the joint fits, we identify three shared resonances at $f_1 = 0.62$ THz, $f_2 = 0.74$ THz, and $f_3 = 0.88$ THz, with corresponding linewidths $\gamma_1 = 81$ GHz, $\gamma_2 = 75$ GHz, and $\gamma_3 = 182$ GHz. In $WS_2$, the mode at $f_2$ carries 59% of the total oscillator strength, followed by 33% from $f_3$, while $f_1$ contributes only 8%. In $WSe_2$, the dominant contribution arises from $f_1$(68%), with $f_2$ and $f_3$ contributing equally (16% each). A summary of fit results for all angle-aligned devices is provided in the Supplementary Information ("Phonon Dynamics Fit Results").

Notably, the effective oscillator strengths can take either sign. For example, $d_2^{WSe_2}$ is consistently negative across all devices, while the other modes ($d_1^{WSe_2}, d_3^{WSe_2}$) are positive.

Similarly, $d_1^{WS_2}$ is negative relative to the remaining modes ($d_2^{WS_2}, d_3^{WS_2}$). The change in the magnitude and sign of $d_k^{(i)}$ is a manifestation of different exciton-phonon physics for different moiré phonon and moiré exciton pairs.

Phenomenologically, the effective oscillator strength $d_k^{(i)}$ encodes two distinct physical contributions, which we write as $d_k^{(i)} = u_k z_k$. The factor $u_k$ describes how efficiently a phonon mode is driven by the arriving phonon wavepacket, while $z_k$ quantifies how strongly a given mode modulates the exciton energy (i.e., the exciton-phonon coupling). In our experiment, excitation originates from out-of-plane lattice motion driven by the FLG transducer, so only modes with finite out-of-plane displacement are efficiently excited. We therefore interpret $u_k$ as the projection of each phonon eigenmode onto this driven coordinate. The phase of the moiré phonon eigenmodes can be chosen so that $u_k$ is always positive. Therefore, within this framework, the relative sign of $d_k^{(i)}$ reflects that of $z_k$, the effective exciton-phonon coupling coefficient. Consequently, a positive (negative) oscillator strength indicates that the corresponding moiré phonon initially redshifts (blueshifts) the moiré exciton resonance.

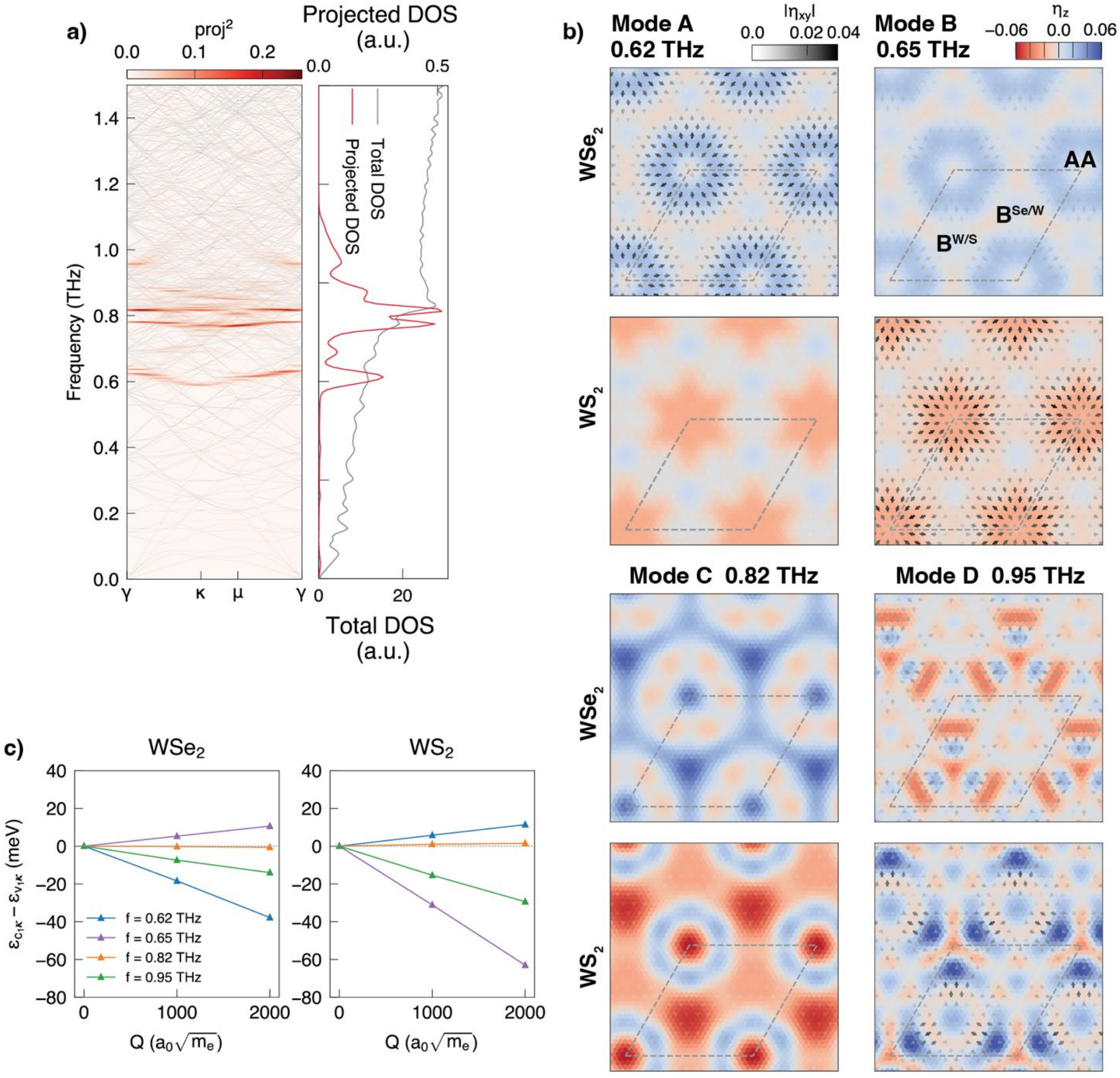

***Figure 4 a)*** *Phonon band structure of angle-aligned* $WS_2/WSe_2$ *heterobilayer. Gray lines denote the phonon dispersion, and the red-color intensities (spectral weights) denote the amplitude square of the projection of the phonon eigenvector to the breathing-mode direction. The right panel displays the total phonon density of states (DOS) (gray) and the breathing-mode-projected DOS (red).* ***b)*** *Normalized polarization vectors* $\vec{\eta}$*, i.e., atomic displacement patterns, of indicated selected zone-center phonon eigenmodes for the W atoms in* $WSe_2$ *(first and third rows) and* $WS_2$ *(second and fourth rows). Arrows indicate the in-plane atomic displacements (*$\eta_{xy}$*), while the blue-to-red color map shows the out-of-plane displacement (breathing-mode) component (*$\eta_z$*), with red (blue) corresponding to motion along the +z (-z) direction.* ***c)*** *Variation of the interband gap at the* $\boldsymbol{\kappa}$ *point under finite-displacement phonon perturbations for 4 selected phonon modes. The left panel shows the gap between the first valence and first conduction bands with dominant* *wavefunction in the $WSe_2$ layer, while the right panel shows the corresponding gap between bands with dominant wavefunction in the $WS_2$ layer.*

To understand the microscopic origin of the observed exciton-phonon coupling, we perform first-principles atomistic simulations of the aligned $WS_2/WSe_2$ heterobilayer. We first compute the moiré phonon spectrum and estimate how efficiently each moiré phonon eigenmode can be driven by the out-of-plane phonon wavepacket. We then calculate how strongly each phonon mode modulates the electronic states and corresponding interband gaps. Together, these two effects microscopically determine the experimentally observed phonon-induced moiré exciton reflectance modulation.

Zone folding from the moiré supercell produces a dense set of moiré phonon modes (Fig. 4a, left panel) in the moiré Brillouin zone. We focus on the subset of moiré phonons that can be efficiently excited by the out-of-plane phonon wavepacket in the experiment. To identify these modes, we project the eigenvector of the moiré phonon eigenmodes onto the interlayer breathing–mode coordinate. The resulting projected density of states (Fig. 4a, right panel) is concentrated in the 0.6–0.9 THz range, in good agreement with the frequency range of the experimentally observed resonances. We identify zone-center moiré phonon modes that carry the largest breathing-mode weight at 0.62 (mode A), 0.65 (mode B), 0.82 (mode C), and 0.95 THz (mode D). Their layer resolved displacement patterns are presented in Fig. 4b, and their key quantitative features are summarized in Table 1

We fix the phase of the 4 relevant moiré phonon eigenmodes so that each mode has a positive out-of-plane displacement at AA site of the $WS_2$ layer. This convention is motivated by the experimental geometry, where the arriving phonon wavepacket is expected to couple most strongly through the points of closest contact between the three-dimensionally reconstructed moiré structure (Fig. S13) and the surrounding hBN. We find that modes A, B and D exhibit strongly hybridized in-plane and out-of-plane vibration character due to the three-dimensional atomic reconstruction of the moiré heterostructure, while mode C remains predominantly layer-breathing in character. This emergent hybridization is a distinct feature of the reconstructed moiré heterostructure.

| Frequency (THz) | $P_B^2$ | $\max\lvert\eta_{xy}^{WSe_2}\rvert$ | $\max\lvert\eta_{xy}^{WS_2}\rvert$ |
|---|---|---|---|
| **A** 0.62 | 0.113 | 0.038 | 0.008 |
| **B** 0.65 | 0.175 | 0.013 | 0.039 |
| **C** 0.82 | 0.282 | 0.007 | 0.010 |
| **D** 0.95 | 0.106 | 0.024 | 0.037 |

***Table 1*** Quantitative summary of selected zone-center moiré phonon modes shown in Fig. 4b, including their squared breathing-mode projection amplitude $P_B^2$ and the maximum magnitudes of the in-plane ($\eta_{xy}$) displacements in the $WSe_2$ and $WS_2$ layers.

We determine the modulation of the moiré exciton transition energy from the individual moiré phonons (the exciton-phonon coupling strength) using density functional theory (DFT) frozen-phonon calculations of the electronic structure. Performing large-scale exciton spectrum calculations for each moiré phonon configuration is computationally demanding. We therefore track the modulation of the interband gaps between the relevant conduction and valence bands as an approximation to the phonon-induced exciton energy shift. Figure 4c shows the calculated band gap changes of the $WSe_2$ and $WS_2$ electronic states as a function of finite phonon-mode displacement. Our calculations reveal that these energy shifts are dominated by the in-plane phonon displacements that modulate the local strain field of the moiré heterostructure. For example, moiré phonon mode C has negligible exciton-phonon coupling (Fig. 4c) because it induces minimal in-plane strain (Fig. 4b). The layer-resolved strain fields associated with each mode are shown in Fig. S15 and discussed in the supplementary information ("Phonon-induced local strain fields"). We also report the diagonal electron-phonon coupling matrix elements for both the $WSe_2$ and $WS_2$ layers in Tables S4 and S5, respectively. These matrix elements are consistent with the trends inferred from the frozen-phonon calculations.

Our first-principles calculation accounts for the major features of the moiré phonon-induced exciton modulation observed in our experiment. First, we assign the three experimentally observed Lorentz resonances $L_1$ (0.62 THz), $L_2$ (0.74 THz) , and $L_3$ (0.88 THz) to the theoretically calculated moiré phonon modes A (0.62 THz), B (0.65 THz), and D (0.82 THz), respectively, as these three phonon modes can (1) be excited efficiently with the out-of-plane phonon wave packet and (2) strongly shift the exciton resonance energy. The calculated frequencies are close to the experimentally observed values, with a small underestimation of the $L_2$ and $L_3$ frequencies.

With this assignment, we examine both the magnitude and sign of the moiré phonon-induced exciton energy shift for different moiré phonon and exciton combinations. The $WSe_2$ Peak I and $WS_2$ moiré exciton are primarily localized near the AA site of the $WSe_2$ layer and B sites of the $WS_2$ layer, respectively. As shown in Fig. 4b, moiré phonon mode A generates strong extensional in-plane strain near the $WSe_2$ AA site, producing a large redshift of the $WSe_2$ Peak I-exciton (Fig. 4c). At the same time, mode A induces only weak compressive strain in the $WS_2$ layer, leading to a small blueshift of the $WS_2$ moiré exciton. In contrast, moiré phonon mode B produces strong extensional in-plane strain in the $WS_2$ layer together with weaker compressive strain in the $WSe_2$ layer AA site, generating a large redshift of the $WS_2$ and a small blueshift of the $WSe_2$ Peak I-exciton. Finally, moiré phonon mode D generates moderate extensional strain near the AA site of the $WSe_2$ layer and B sites of the $WS_2$ layer, leading to moderate redshifts of the moiré excitons in both layers.

Overall, the first-principles calculated signs and relative magnitudes of the exciton energy shifts can account for the experimentally observed moiré exciton-phonon coupling, supporting the assignment of resonances $L_1$, $L_2$ and $L_3$ to moiré phonon modes A, B and D.

**Conclusion and Outlook**

In summary, we combine ultrafast terahertz phonon spectroscopy and *ab initio* calculations to elucidate the unusual moiré phonon- moiré exciton coupling in $WS_2/WSe_2$ heterobilayers. We show that moiré phonons with hybrid in-plane and out-of-plane vibration can emerge from the three dimensionally reconstructed moiré heterostructures, and each of these moiré phonons can couple differently to various moiré excitons in the $WS_2$ and $WSe_2$ layers. Our approach provides a general route to resolve and ultimately control mode-specific exciton–phonon interactions in van der Waals heterostructures, with potential applications in ultrafast modulation, TMD-based nanoacoustic cavities[34,36,41,42] and phonon-engineered optoelectronics. More broadly, these results establish moiré exciton–phonon coupling as a new avenue for engineering non-equilibrium optical properties in quantum materials through targeted control of lattice dynamics[43].

## Methods

*Device Fabrication and Characterization:* Van der Waals heterostructures were fabricated using well-established tear-and-stack techniques. Bulk crystals of $WSe_2$, $WS_2$, hBN, and few-layer graphene (FLG) were mechanically exfoliated onto $SiO_2$/Si substrates. Monolayer flakes of $WSe_2$ and $WS_2$, along with hBN flakes of the desired thickness and suitable FLG flakes, were identified based on their optical reflection contrast under white-light illumination.

Selected flakes were sequentially picked up and vertically assembled into the target heterostructure using a polycarbonate stamp and subsequently released onto the desired substrate. Detailed descriptions of the assembled heterostructures and their corresponding substrates are provided in the Supplementary Information.

.

*Ultrafast Pump Probe Spectroscopy:* We used an amplified Yb:KGW femtosecond laser (Light Conversion Carbide) operating at a repetition rate of 151 kHz to pump a commercial optical parametric amplifier (OPA, Light Conversion Orpheus Twins). The Carbide produced pulses centered at 1030 nm with a duration of 200 fs and a pulse energy of 52.9 μJ. The signal output of the OPA was tuned to 880 nm and used as the pump beam for all experiments.

The residual 1030 nm beam following the OPA was focused onto a 3 mm thick sapphire plate to generate a supercontinuum. This supercontinuum was spectrally filtered to the 550–800 nm range and used as the probe beam. Both pump and probe beams were vertically polarized for all experiments described in this manuscript.

The pump and probe beams were combined using a beamsplitter and focused onto the sample with a 20× objective lens (NA = 0.45). The reflected probe light was collected, spectrally dispersed by a transmission grating (300 grooves/mm), and imaged onto a linear array detector (Coptonix S11639-01) using a lens with a focal length of 5 cm.

The pump beam was modulated at 707 Hz using an optical chopper. Reflected pump light was suppressed with a short-pass filter (Thorlabs FESH850). The pump–probe reflection contrast spectra at the modulation frequency at each time delay were averaged over tens of thousands of laser shots to achieve a satisfactory signal-to-noise ratio. Group delay dispersion introduced into the supercontinuum probe was corrected during post-processing.

*First-principles atomistic and electronic structure simulation*: To obtain the equilibrium structure and phonon spectrum of the angle-aligned $WS_2/WSe_2$ heterobilayer, we constructed a moiré supercell of 25 × 25 $WSe_2$ and 26 × 26 $WS_2$, relaxed its atomic structure, and computed the phonons using a finite-displacement approach. The resulting force-constant matrix was used to evaluate the phonon density of states and phonon dispersion in the moiré Brillouin zone. To identify the phonon modes most relevant to the experiment, we projected the calculated moiré phonon eigenvectors onto an idealized zero-in-plane-momentum interlayer breathing pattern and used the squared breathing-mode projection weight to isolate the moiré phonon modes with the strongest overlap with the experimentally driven interlayer breathing coordinate. To assess how these phonons perturb the electronic structure, we performed frozen-phonon first-principles calculations based on the calculated phonon eigenvectors and extracted the resulting changes in the relevant interband gaps and band-edge eigenvalues. Additional details of the structural relaxation, phonon calculations, breathing-mode projection analysis, and frozen-phonon electronic-structure calculations are provided in the Supplementary Information.

## Acknowledgements

This work was primarily supported by the US Department of Energy, Office of Science, Basic Energy Sciences, Materials Sciences and Engineering Division (DE-AC02-05CH11231 within the Nanomachine Program), including pump–probe spectroscopy, data analysis and atomistic and electronic structure simulations. The monolayer exfoliation and heterostructure stacking were supported by the US Department of Energy, Office of Science, Basic Energy Sciences, Materials Sciences and Engineering Division (DE-AC02-05CH11231 within the van der Waals Heterostructure Program, KCWF16). C.U. acknowledges support from the Kavli ENSI Graduate Research Fellowship. We acknowledge the Texas Advanced Computing Center (TACC) at the University of Texas at Austin for providing high-performance computing resources. This research also used resources of the National Energy Research Scientific Computing Center (NERSC), a US Department of Energy, Office of Science User Facility, located at Lawrence Berkeley National Laboratory, operated under contract no. DE-AC02-05CH11231. M.H.N. acknowledges support from the National Science Foundation through the Center for Dynamics and Control of Materials:

an NSF MRSEC under Cooperative Agreement No. DMR-2308817. M.J. acknowledges the Nano mission of the Department of Science and Technology, India, and the Anusandhan National Research Foundation, India, for financial support under Grants No. DST/NM/TUE/QM-10/2019 and No. ANRF/ARG/2025/003370/PS, respectively. K.W. and T.T. acknowledge support from the CREST (JPMJCR24A5), JST and World Premier International Research Center Initiative (WPI), MEXT, Japan.

# Supplementary Information to Moiré Phonons and Emergent Exciton-Phonon Coupling in a Moiré Heterobilayer

## Monolayer vs Moiré in Time Domain:

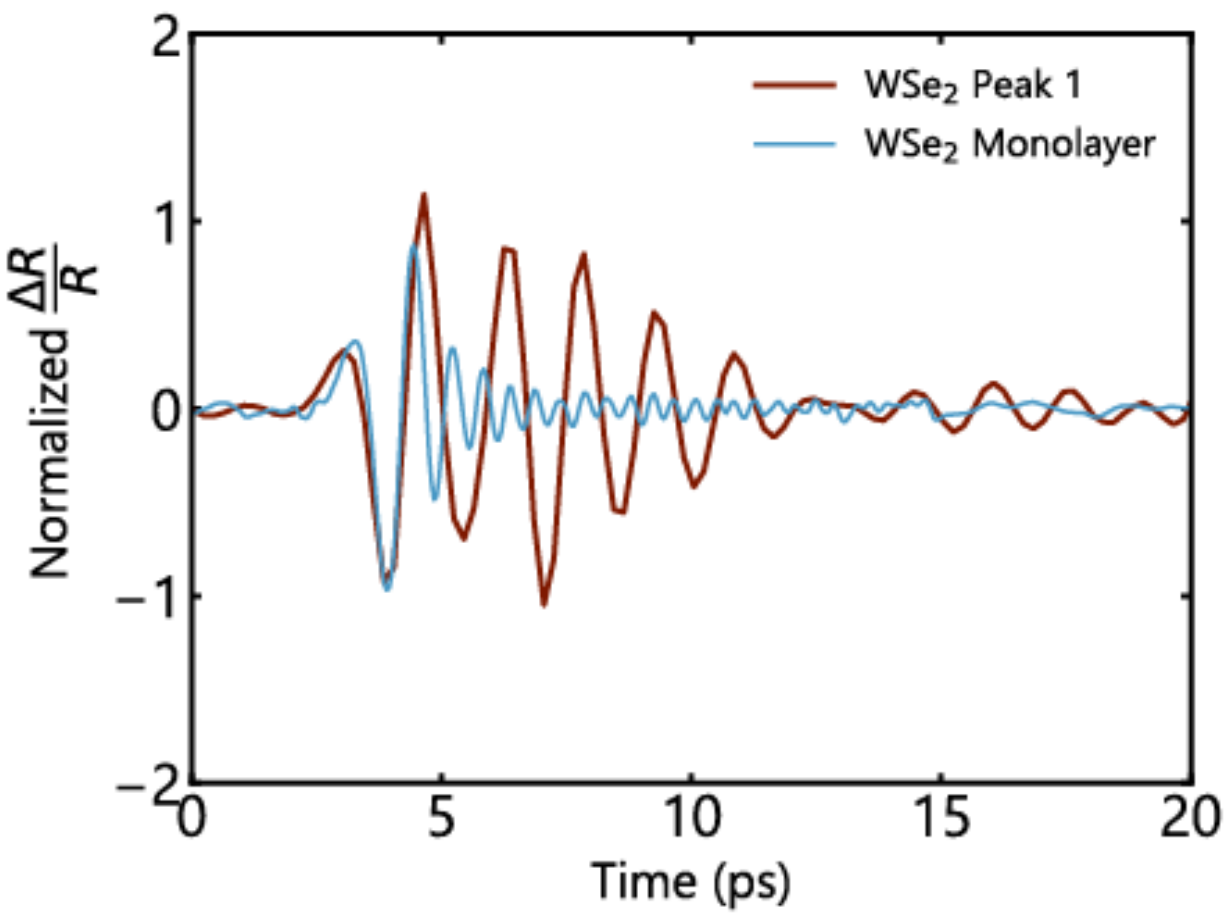

***Figure S1*** *Terahertz phonon response measured from $WSe_2$ Peak I exciton in an aligned $WS_2/WSe_2$ sample overlaid with the response measured from $WSe_2$ exciton from a monolayer sample.*

## Device Parameters:

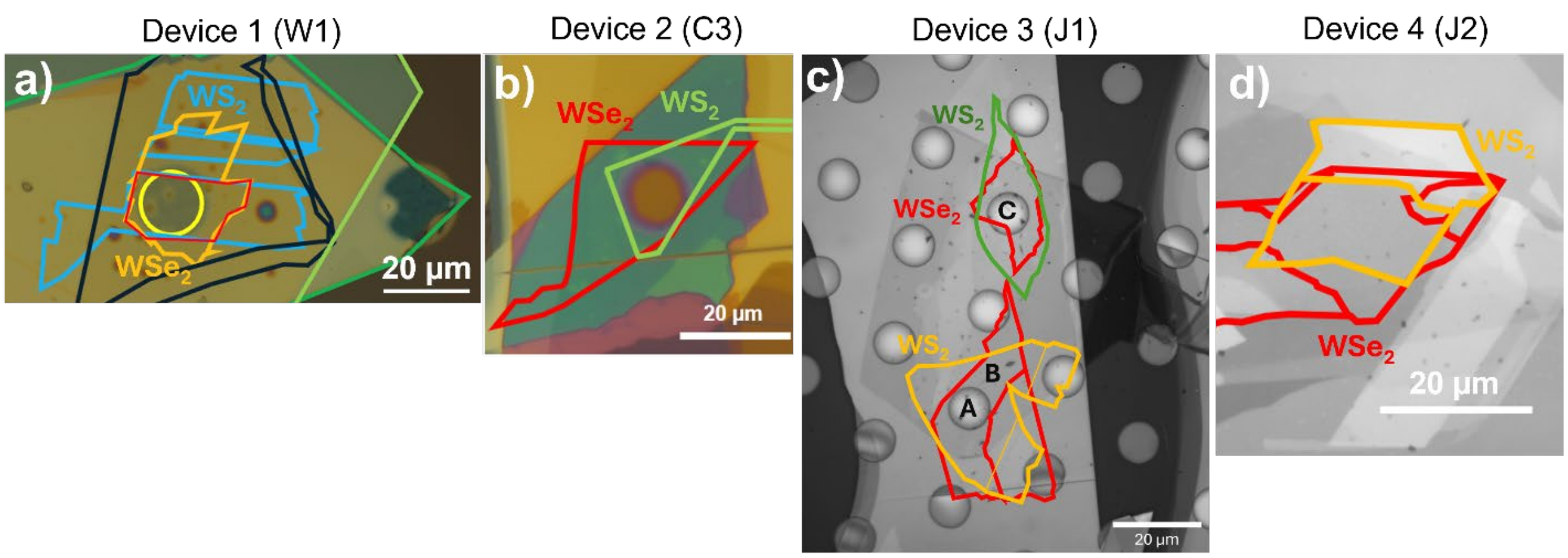

***Figure S2*** *Microscope images of all the $WS_2/WSe_2$ devices measured in this study. In total 4 different devices were made.*

We reproduced the emergent phonon dynamics reported in the main text across four distinct devices. Optical microscopy images of these devices are shown in Figure S2.

**Device 1** consists of FLG / 30 nm hBN / angle-aligned $WS_2/WSe_2$/ 50nm hBN van der Waals heterostructure stack. The substrate is 90nm $SiO_2$ on bulk Si, with a donut shaped 50nm Au layer deposited on top of the $SiO_2$. The heterostructure is released onto the Au ring such that the entire stack is suspended over an air gap. This geometry was designed to minimize the loss of the reflected phonon packet upon its return to the TMD; if the stack were placed directly on the 90nm $SiO_2$substrate, a portion of the reflected phonon would transmit into the substrate. For the purposes of the present work, however, the air gap is not essential and in subsequent devices we removed the air gap feature from our device design.

**Device 2** consists of FLG/ 56 nm hBN/ aligned $WS_2/WSe_2$/50 nm hBN stack released on a 90nm $SiO_2$/Si substrate. In this case, 10 µm diameter holes were pre-etched into the $SiO_2$layer and the van der Waals stack is suspended over these etched regions.

**Device 3** contains three regions of interest, labeled A, B and C. Regions A and B both consist of FLG/45 nm hBN/ aligned $WS_2/WSe_2$/ 50 nm hBN, differing only in their substrate configuration. Region A is suspended over a 90 nm air gap, while Region B is released directly onto the 90 nm $SiO_2$substrate. This design enables a direct comparison of the influence of local optical fields on the measured phonon dynamics and whether the air gap substantially effects the measurements. Region C consists of FLG/45 nm hbN/ angle-misaligned $WS_2/WSe_2$/ 50 nm hBN suspended over a 90 nm air gap-Si substrate.

**Device 4** consists of FLG/42 nm hBN/aligned $WS_2/WSe_2$/50 nm hBN stack released on a 285nm $SiO_2$/Si substrate.

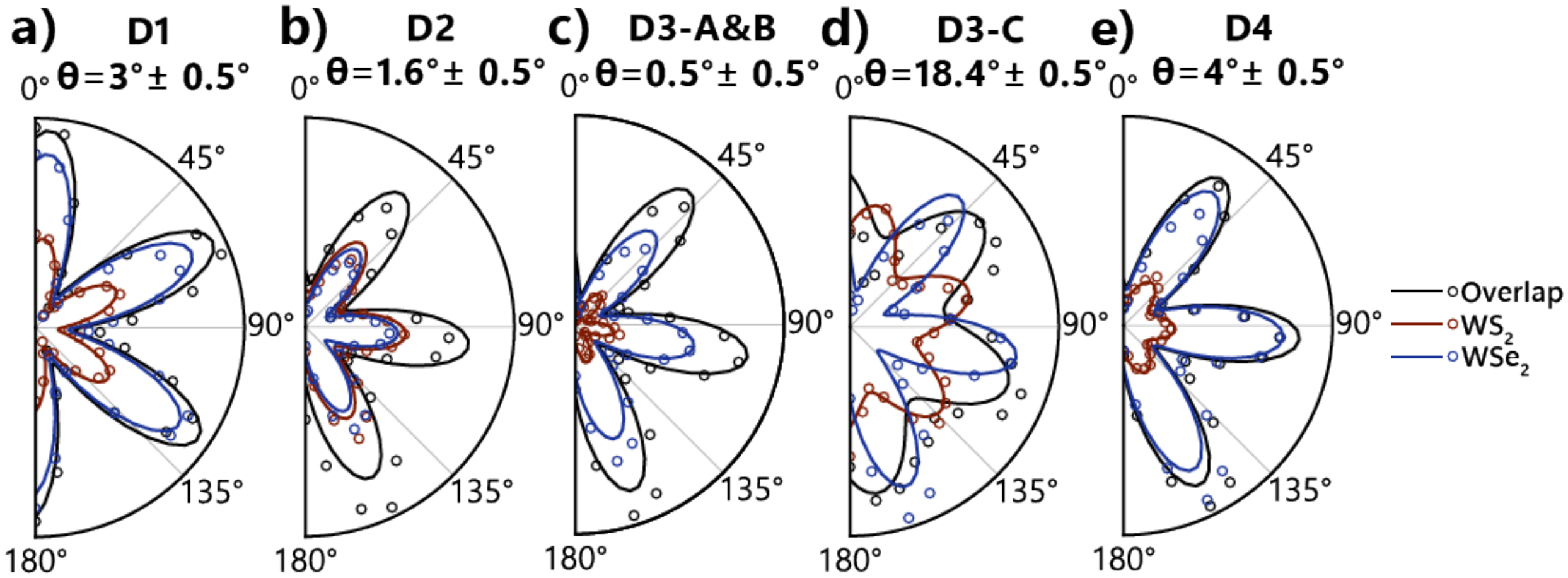

***Figure S3****- Second Harmonic Generation anisotropy of the devices under consideration.*

We use the idler output of the Orpheus Twins OPA, tuned to 1060 nm, as the fundamental wavelength to probe second harmonic generation (SHG) and determine the relative twist angle of the devices. The fundamental beam is focused onto the sample using a 50× microscope objective, and the reflected second harmonic at 530 nm is collected with a cooled camera (Princeton Instruments, PIXIS-100). The residual fundamental is suppressed using a stack of short-pass filters (Thorlabs, FESH850). A half-wave plate and a polarizer are inserted into the beam path to simultaneously vary the incident and detected polarizations, allowing us to record the SHG intensity as a function of polarization. This yields the expected sixfold symmetric pattern (Fig. S3). By scanning the excitation spot across individual TMD regions and their overlap, we extract the twist angle (θ) from the relative orientation of the SHG patterns, these are reported in Fig. S3 for each device. The uncertainty in the assigned twist angle is extracted from the fit residuals.

**Fabry–Pérot Effects:**

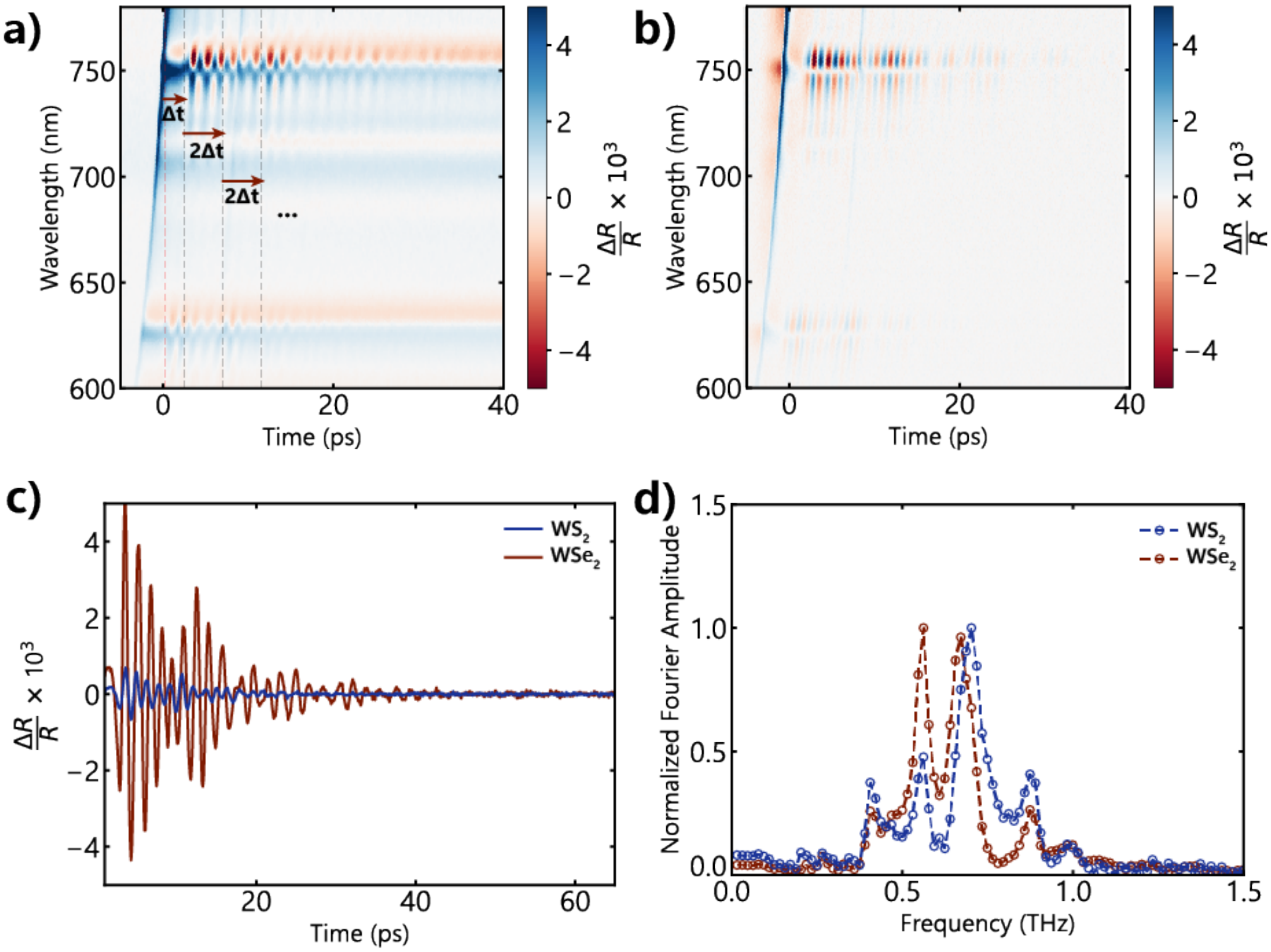

***Figure S4*** *Terahertz phonon response measured from a moiré $WS_2/WSe_2$ sample but with a thin b-hBN between the FLG and the TMDs.* ***a)*** *The raw reflection contrast spectrum shows the arrival of the initial phonon within 2.9ps of the initial time, followed by repeated revivals marked by arrows* ***b)*** *Phonon dynamics after filtering out the slow responses as described in the main text* ***c)*** *Phonon time traces reported by $WSe_2$ peak I and $WS_2$ A excitons and* ***d)*** *their corresponding Fourier amplitudes.*

We stress the importance of using thick hBN layers between the phonon transducer and the TMD to extract the intrinsic phonon dynamics of the TMD. If the hBN layers are too thin, the dynamics induced by the initial phonon packet interfere with multiple reflections of the back-reflected phonon packets at later times, effectively forming a phonon Fabry-Perot cavity

Experimentally, the initial phonon packet persists for ~15 ps in angle-aligned samples. Therefore, the round-trip time of the transmitted and reflected phonons must exceed this duration to avoid overlapping with the initial response. Taking the speed of sound in hBN to be 3.5 km/s, this requirement corresponds to a minimum hBN thickness of 26.25 nm for both top and bottom hBN layers. In all data presented in the main text, we use devices with the hBN thicknesses larger than this lower bound.

To illustrate how Fabry-Perot resonances obscure the intrinsic material response, we present phonon dynamics measured in a moiré $WS_2/WSe_2$ sample with a thin b-hBN separating the FLG from the TMDs. In this case, the initial phonon pulse reaches the TMD within ~2.9 ps, as marked by the vertical dashed lines in Figure S4a, corresponding to hBN thickness of ~10 nm. This is followed by multiple revivals of the phonon signal at twice this time delay, corresponding to the reflected phonon packet completing a round trip and returning to the TMD.

Figure S4c shows the phonon time traces measured via the $WSe_2$ Peak I exciton and the $WS_2$ A exciton. Both traces differ from those presented in the main text. In Figure S4d, we plot the amplitude spectra of these time traces in frequency domain. Instead of the response shown in Figure 3 of the main text, we observe multiple sharp peaks for both $WSe_2$ and $WS_2$. As described above, this is the result of a Fabry-Perot cavity effect. On average the experimentally observed free spectral range (FSR) is 0.15 THz consistent with the expected FSR of 0.17 THz for a phonon cavity with a 10nm hBN spacer ($\Delta f = \frac{v_s}{2d}$).

With prior knowledge of the intrinsic response of the moiré $WS_2/WSe_2$, one can interpret the Fabry-Perot spectrum and attribute the slight shift in the central frequency of the third $WS_2$ peak relative to the third $WSe_2$ Fabry-Perot peak to the intrinsically higher resonance frequency of $WS_2$ compared to $WSe_2$. However, without such prior knowledge, this distinction is not readily apparent.

**Time Domain Comparison:**

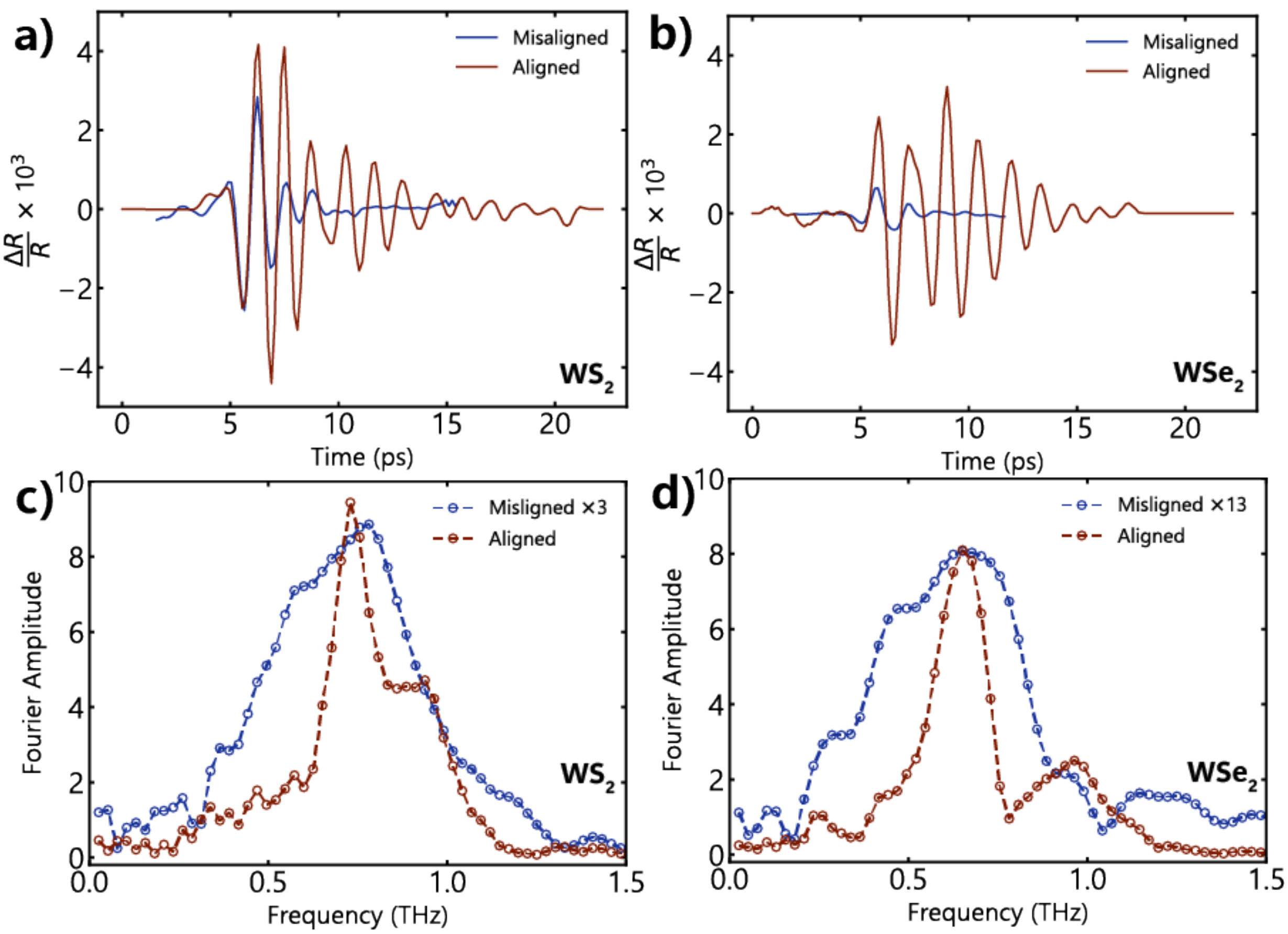

***Figure S5** Direct comparison between misaligned $WS_2/WSe_2$ and aligned $WS_2/WSe_2$ from the same sample **a)** Phonon dynamics measured through $WS_2$ exciton. **b)** Phonon dynamics measured through $WSe_2$ exciton and their corresponding fourier amplitude spectra for **c)** $WS_2$ and **d)** $WSe_2$.*

In addition to the enhanced phonon lifetime observed in angle-aligned $WS_2/WSe_2$ compared to the angle misaligned case, we also observe an apparent increase in the amplitude of the phonon response for the angle-aligned $WS_2/WSe_2$.

The absolute amplitude of the phonon signal is sensitive to sample-to-sample variations. These variations primarily arise from differences in the optical local field at both the FLG phonon transducer and the TMD layers, which depend on the substrate and specific hBN thicknesses used. Additional variability can result from differences in interface quality introduced during the repeated tear and stack fabrication process, further influencing the amplitude of the measured phonon transients.

To control these external factors, we fabricated a FLG/ 45 nm hBN/ $WS_2$-$WSe_2$/ 50 nm hBN sample containing two distinct $WS_2/WSe_2$ regions. One region consisted of an angle-aligned and the other angle misaligned $WS_2/WSe_2$. In Fig S5a and b, we present the absolute amplitude of the phonon induced dynamics as measured via the $WS_2$ and $WSe_2$ excitons, respectively. The enhancement of the phonon amplitude is particularly pronounced in the $WSe_2$ exciton response. Fig S5c and d show the corresponding Fourier amplitude spectra for the $WS_2$ and $WSe_2$ excitons. The Fourier amplitude of the angle-aligned $WSe_2$ spectrum is approximately 13 times larger than that of the angle-misaligned case, while for the $WS_2$ the enhancement factor is approximately 3.

**Natural Bilayer:**

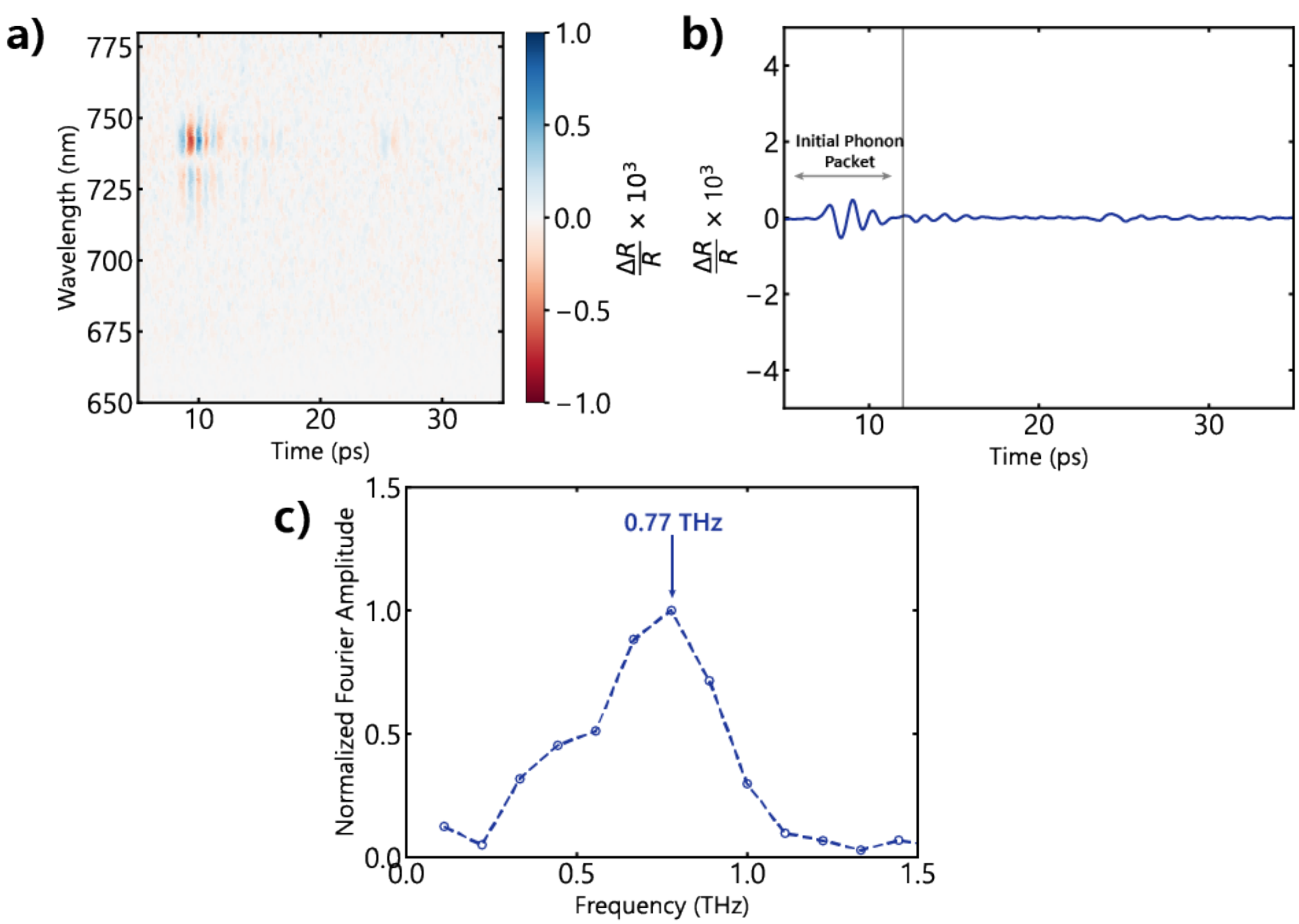

***Figure S6*** *Terahertz phonon response measured from a natural bilayer $WSe_2$ sample.* ***a)*** *Phonon dynamics after filtering out the slow responses as described in the main text.* ***b)*** *Phonon time traces reported by $WSe_2$ exciton and* ***c)*** *the corresponding Fourier amplitude spectrum of the initial phonon dynamics.*

To clarify the origin of the long-lived phonon dynamics observed in the moiré $WS_2/WSe_2$ samples, Figure S6a presents the phonon measurements performed on a natural bilayer $WSe_2$ sample. The sample structure is FLG / 27 nm hBN / bilayer $WSe_2$ / 8 nm hBN.

In this case, the phonon dynamics are short lived which enables a clear temporal separation between the initial phonon response and the subsequent revival of the signal arising from the round-trip propagation of the transmitted phonon through the thin top hBN layer, as shown in Fig. S6b.

Figure S6c displays the Fourier transform of the initial phonon packet. The spectrum exhibits a broad feature centered around 0.77 THz, closely resembling the angle-misaligned response and distinctly different from the moiré response.

These observations support the conclusion that the long-lived phonon dynamics are a consequence of the moiré-modified lattice and the emergence of moiré phonon modes, rather than a simple modification of the phonon modes in a conventional bilayer structure.

**Phonon Model:**

We implement a phenomenological model to fit the observed time domain traces using an effective susceptibility, $\chi_{eff}(\omega)$. In this model, a phonon wavepacket with a Gaussian envelope $U(\omega)$ drives three Lorentzian oscillators $L_k$. The complex frequency-domain response is given by

$$H^{(i)}_{model}(\omega) = \chi^{(i)}_{eff}(\omega)\, U(\omega)$$

which is transformed to the time domain via an inverse Fourier transform

$$H^{(i)}_{model}(t) = \mathcal{F}^{-1}\left[H^{(i)}_{model}(\omega)\right]$$

where $i = WS_2$ or $WSe_2$.

We model the effective susceptibility as,

$$\chi^{(i)}_{eff}(\omega) = \sum_{k=1}^{3} L^{(i)}_k(\omega)$$

where each phonon mode is modeled as a Lorentzian oscillator of the form

$$L^{(i)}_k(\omega) = \frac{d^{(i)}_k\, \omega_k^2}{\omega_k^2 - \omega^2 - i\gamma_k\omega}$$

Here, $d^{(i)}_k$ is the oscillator strength of the layer $i$, $\omega_k = 2\pi f_k$ is the resonance frequency and $\gamma_k$ is the linewidth.

We employ a joint fitting routine that minimizes a least squares cost function simultaneously for both $WS_2$ and $WSe_2$ time domain data. Because the phonon modes originate from a common moiré superlattice, their resonance frequencies $\omega_k$ and linewidths $\gamma_k$ are expected to be identical across the two layers. In contrast, the coupling strength to each exciton is layer-dependent. Accordingly, we constrain $\omega_k$ and $\gamma_k$ to be shared parameters, while allowing the oscillator strengths $d_k$ to vary independently for each dataset.

The joint fit therefore contains 12 free parameters: three shared frequencies $\omega_k$, three shared linewidths $\gamma_k$, and two independent sets of oscillator strengths $\{d_k^{WS_2}\}$ and $\{d_k^{WSe_2}\}$. The cost function is constructed as the sum of squared residuals over both datasets,

$$y = \sum_{i=WS_2,WSe_2} \sum_t (H_{measured}^{(i)}(t) - H_{model}^{(i)}(t))^2$$

which we minimize using a heuristic global optimizer (basin hopping) with randomized initial conditions.

The driving phonon wavepacket is modeled as:

$$U(\omega) = e^{-\left(\frac{(\omega-\omega_c)^2}{2\sigma^2}\right)} e^{j\phi(\omega)}$$

To describe the Gaussian envelope of the driving phonon packet, we use $\omega_c = 2\pi \times 0.78$ THz and $\sigma = 2\pi \times 0.26$ THz .

The second exponential term in the drive describes the frequency dependent phase accumulated by the broadband phonon pulse after propagating through hBN. Due to dispersion in hBN, different frequency components of the pulse acquire different phase delays, which in turn affects how the pulse coherently drives dynamics in the $WS_2/WSe_2$ layers.

We model this phase accumulation using the known dispersion of the speed of sound in hBN. The out of plane acoustic velocity is $\nu_s = 3.5$ km/s below 0.6 THz and decreases to about 3 km/s near 2 THz. The phase accumulated by each frequency component after propagating a distance L is given by

$$\phi(\omega) = k(\omega)L$$

where we model the dispersion relation as a piecewise function, above $\omega_i$

$$k(\omega) = \frac{\omega}{\nu_s} + \beta(\omega - \omega_i)^2$$

and below $\omega_i$

$$k(\omega) = \frac{\omega}{\nu_s}$$

Here $\omega_i = 2\pi \times 0.6$ THz is the inflection point in the dispersion of hBN, $\nu_s$ is the speed of sound below $\omega_i$. The parameter $\beta$ controls the strength of the dispersion and is related to the group velocity dispersion (GVD)

$$\frac{d^2k}{d\omega^2} = 2\beta$$

We use $2\beta = 2 \times 10^{-17}\ \mathrm{s^2/m}$ as the dispersion parameter and calculate $\mathrm{k}(\omega)$. We then extract the phase velocity as $\nu_{\mathrm{p}}(\omega) = \frac{\omega}{\mathrm{k}(\omega)}$. Figure S7a shows the extracted phase velocity meanwhile Fig. S7b shows the accumulated phase difference as a function of frequency after removing the trivial linear background. This is essentially the phase background present in all our measurements.

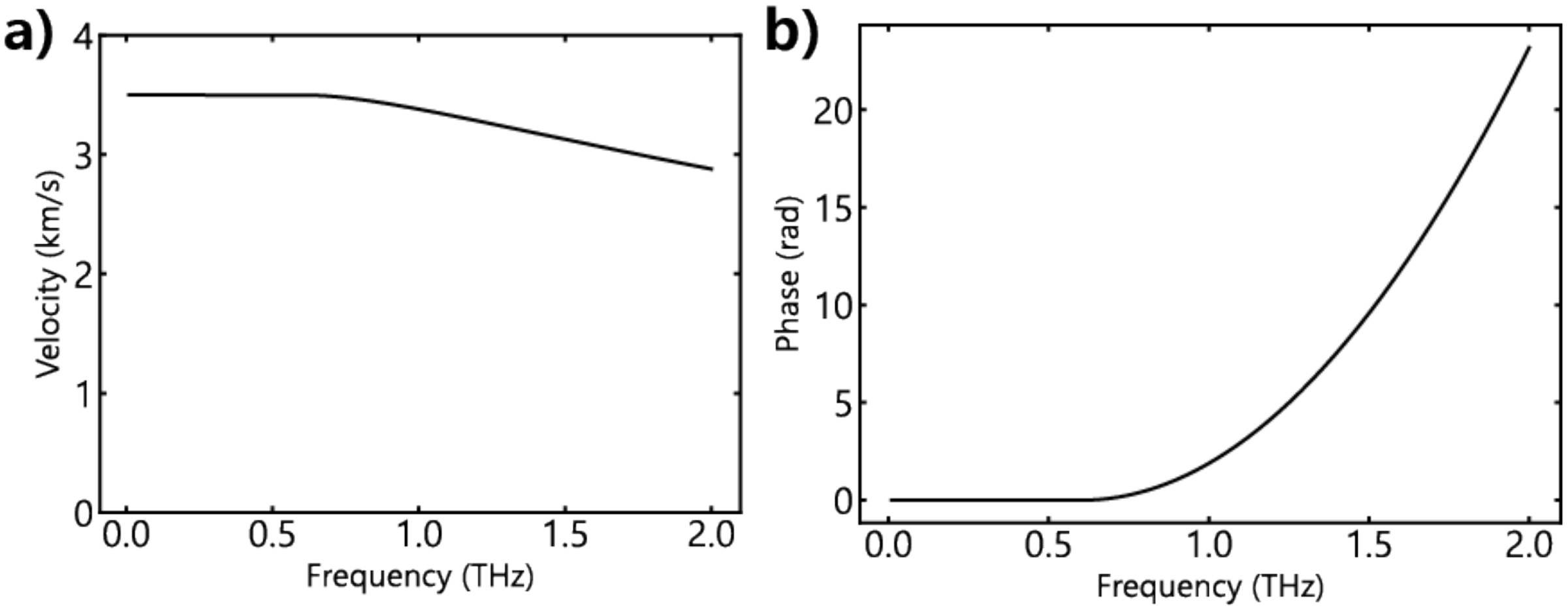

***Figure S8 a)*** *Calculated speed of sound using $2\beta = 2 \times 10^{-17}\ s^2/m$ which is the value we use for the phonon model presented in the manuscript.* ***b)*** *The accumulated phase after traveling $L = 30\ nm$, the linear part of the phase is removed.*

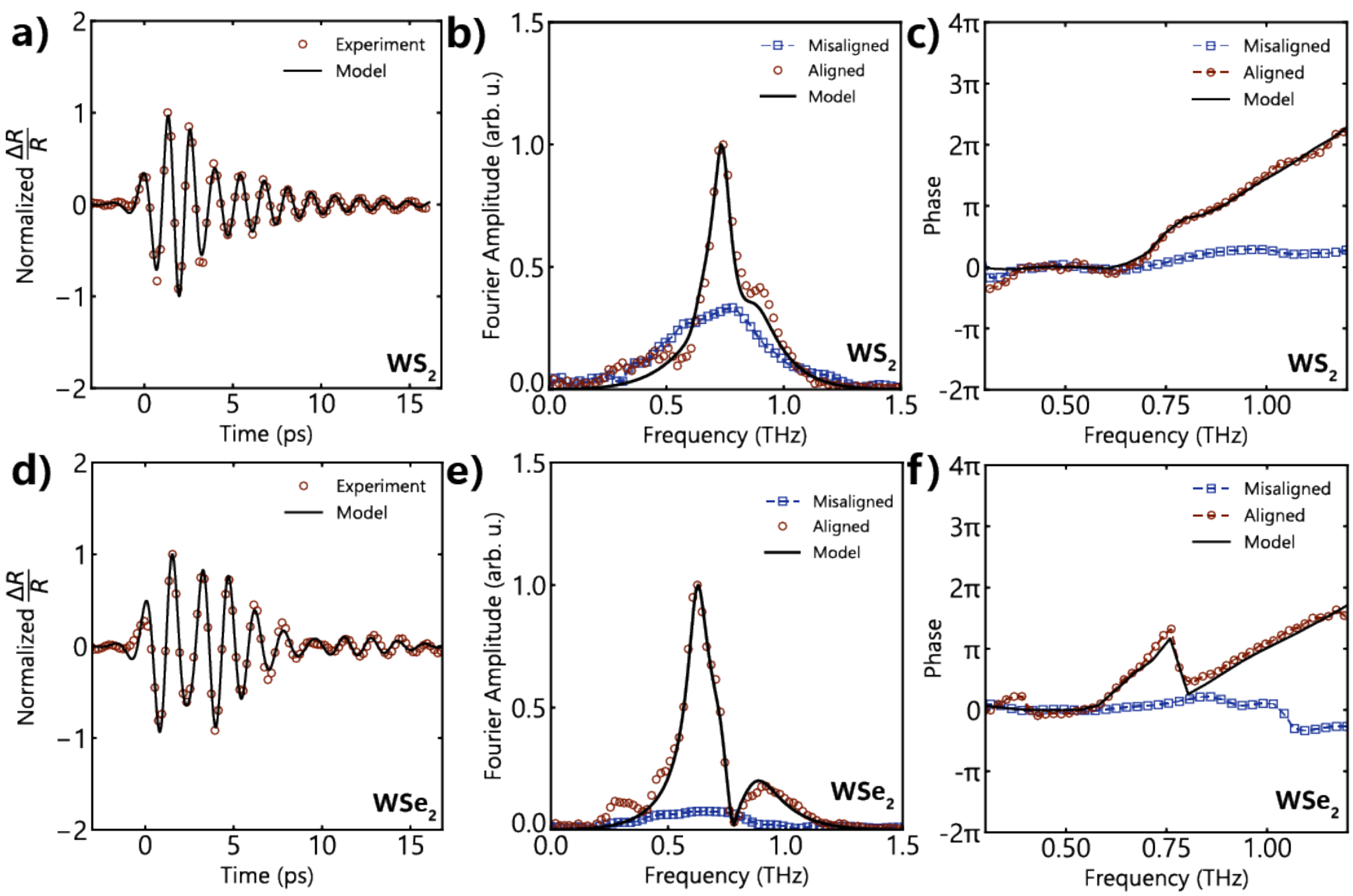

***Figure S8 a)*** *Phonon-induced exciton dynamics of the $WS_2$ A, the black line shows the driven oscillator model fit.* ***b)*** *Corresponding Fourier amplitude spectrum, compared to the large twist angle (misaligned) $WS_2$ response (main text, Fig. 2c).* ***c)*** *Frequency dependent relative phase of the $WS_2$ phonon response.* ***d)*** *Phonon-induced exciton dynamics of the $WSe_2$ Peak I, the black line shows the model fit.* ***e)*** *Fourier amplitude spectrum compared to the large twist angle $WSe_2$ response (main text, Fig. 2c).* ***f)*** *Frequency dependent relative phase of the $WSe_2$ phonon response.*

## Phonon Dynamics Fit Results:

We repeat the above-described fitting routine for four angle aligned devices. Table S1 describes the frequencies and widths of the resonances while Table S2 presents the oscillator strengths extracted from these joint fits. Further Fig. S9 through 11 show the time domain traces, amplitude and phase spectra along with the fits that yield the values in these tables.

| | $f_1$ (THz) | $\gamma_1$ (GHz) | $f_2$ (THz) | $\gamma_2$ (GHz) | $f_3$ (THz) | $\gamma_3$ (GHz) |
|---|---|---|---|---|---|---|
| *Device 1* | 0.62 | 81.58 | 0.74 | 75.2 | 0.88 | 181.8 |
| *Device 2* | 0.63 | 98.3 | 0.75 | 107.6 | 0.88 | 200 |
| *Device 3* | 0.63 | 249.66 | 0.74 | 80.1 | 0.91 | 73.8 |
| *Device 4* | 0.65 | 111.81 | 0.76 | 76.7 | 0.95 | 212.0 |

***Table S1*** *Frequency and widths of the three resonances extracted from four different devices.*

| | $d_1^{WSe_2}$ | $-d_2^{WSe_2}$ | $d_3^{WSe_2}$ | $-d_1^{WS_2}$ | $d_2^{WS_2}$ | $d_3^{WS_2}$ |
|---|---|---|---|---|---|---|
| *Device 1* | 0.68 | 0.16 | 0.16 | 0.08 | 0.59 | 0.33 |
| *Device 2* | 0.68 | 0.2 | 0.12 | 0 | 1 | 0 |
| *Device 3* | 0.53 | 0.21 | 0.26 | 0.13 | 0.39 | 0.48 |
| *Device 4* | 0.59 | 0.2 | 0.21 | 0.03 | 0.4 | 0.57 |

***Table S2*** *Summary of results across the four devices. The columns indicate the oscillator strength of each of the resonances. For each device, the oscillator strength is defined as relative, such that their absolute values sum up to 1 for $WS_2$ and $WSe_2$. $d_2$ is negative for $WSe_2$ meanwhile $d_1$ is negative for $WS_2$.*

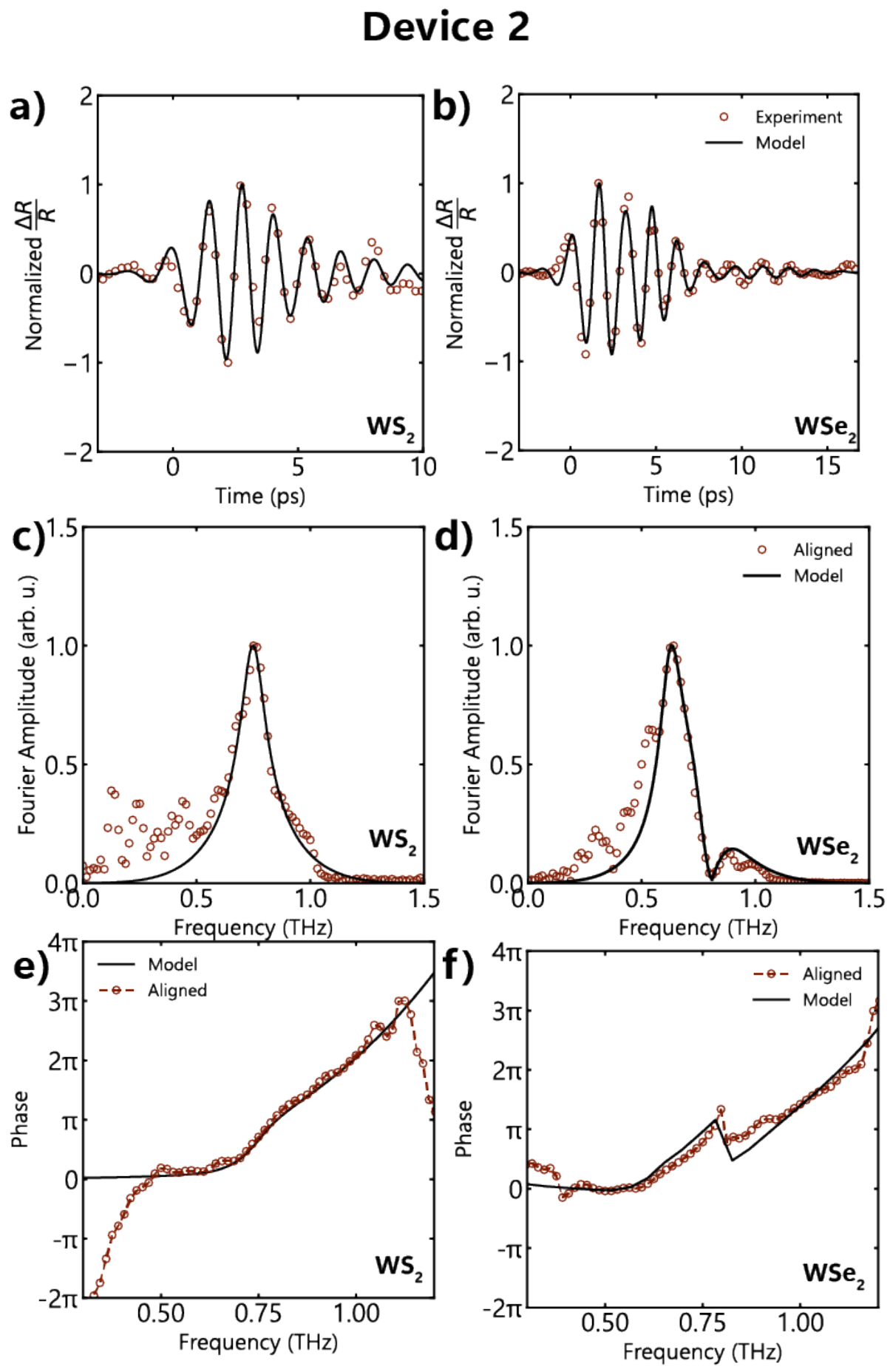

***Figure S9*** *Summary of experiments on Device 2. Time domain trace of* ***a)*** *$WS_2$ and* ***b)*** *$WSe_2$ along with model fits shown in black. Amplitude spectrum for* ***c)*** *$WS_2$ and* ***d)*** *$WSe_2$. Phase spectrum along with the best fit line for L=56nm* ***e)*** *$WS_2$ and* ***f)*** *$WSe_2$.*

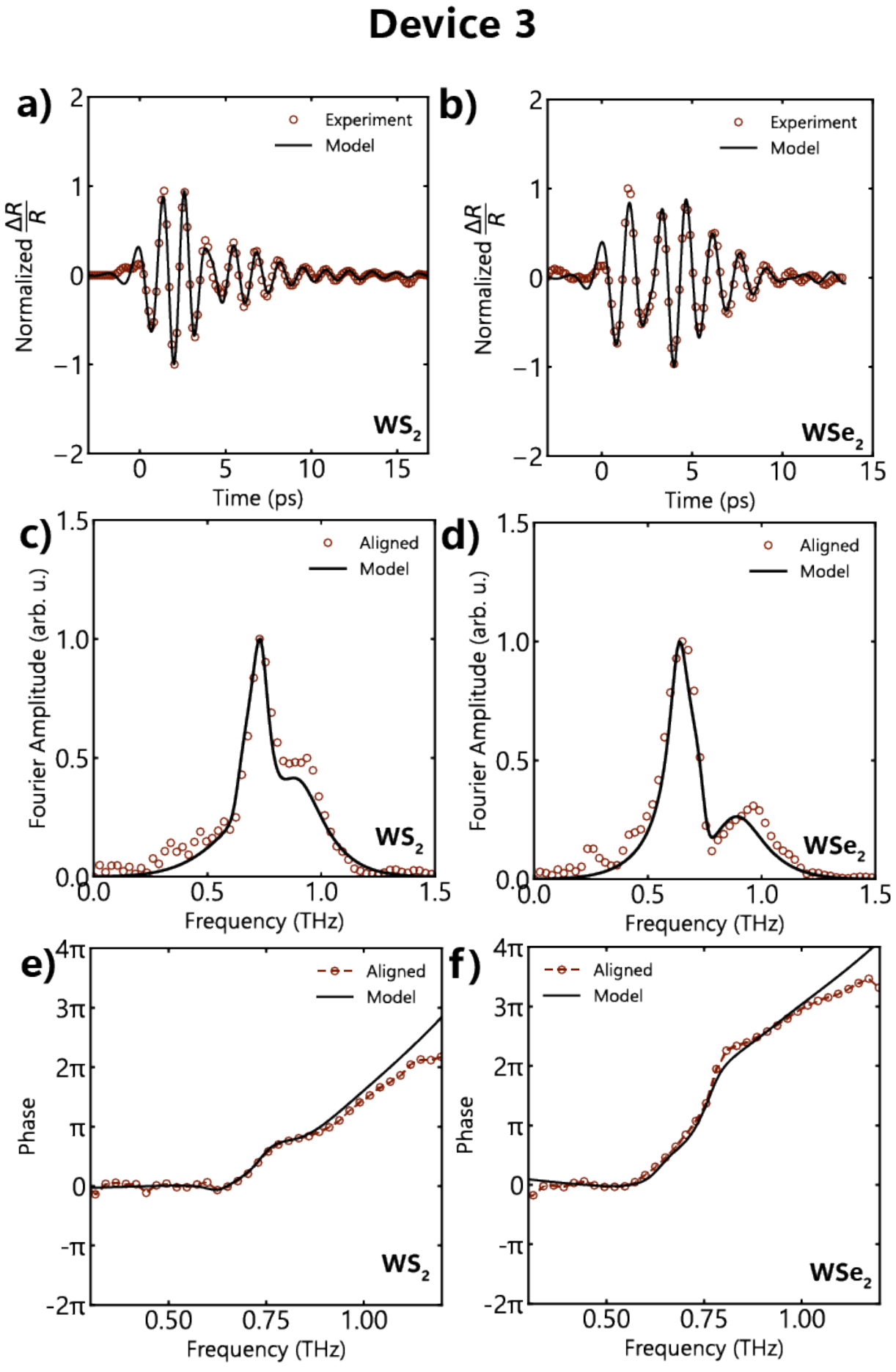

***Figure S10*** *Summary of experiments on Device 3. Time domain trace of* ***a)*** *$WS_2$ and* ***b)*** *$WSe_2$ along with model fits shown in black. Amplitude spectrum for* ***c)*** *$WS_2$ and* ***d)*** *$WSe_2$. Phase spectrum along with the best fit line for L=45nm* ***e)*** *$WS_2$ and* ***f)*** *$WSe_2$.*

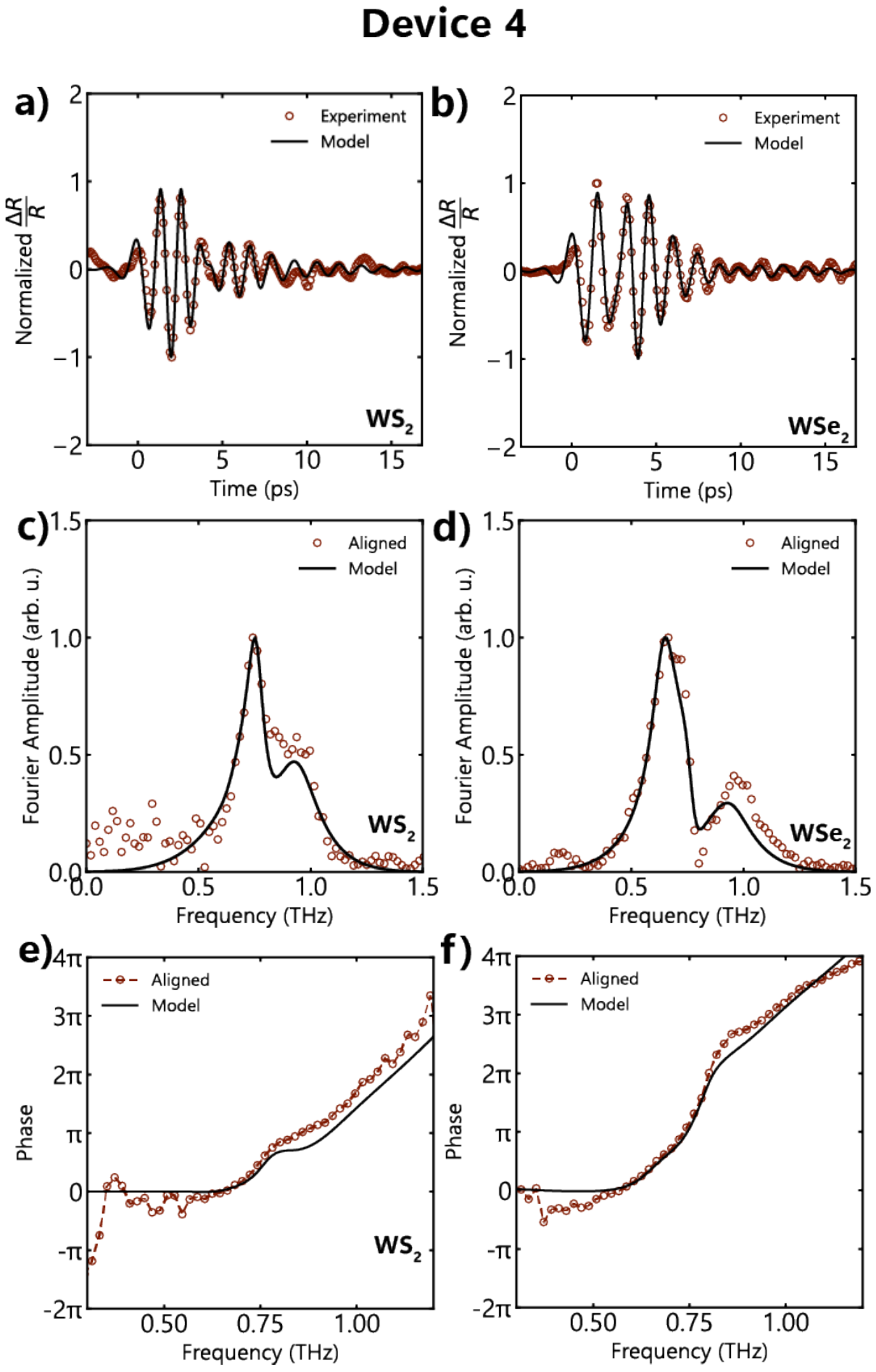

***Figure S11*** *Summary of experiments on Device 4. Time domain trace of* ***a)*** *$WS_2$ and* ***b)*** *$WSe_2$ along with model fits shown in black. Amplitude spectrum for* ***c)*** *$WS_2$ and* ***d)*** *$WSe_2$. Phase spectrum along with the best fit line for L=42nm* ***e)*** *$WS_2$ and* ***f)*** *$WSe_2$.*

**Peak III Dynamics**

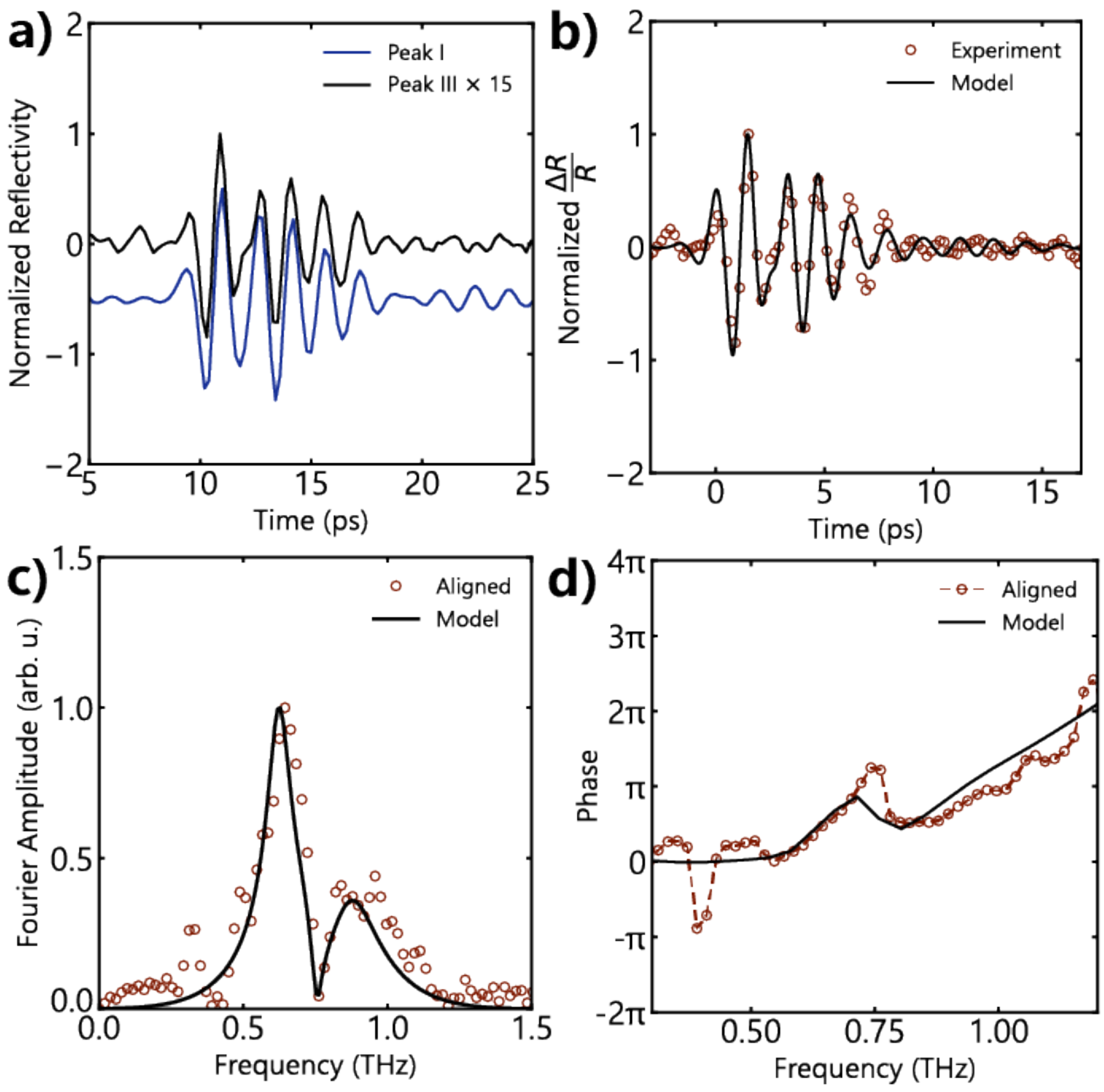

***Figure S12 a)*** *Comparison of phonon-induced exciton dynamics for the $WSe_2$ Peak I and Peak III excitons measured on Device 1.* ***b)*** *Driven-oscillator model fit (black line) to the Peak III exciton dynamics with resonance frequencies and linewidths fixed to the values reported in the main text, while the oscillator strengths are allowed to vary.* ***c)*** *Fourier amplitude spectrum of the Peak III phonon response overlaid with the driven-oscillator fit (black line). d) Phase of the phonon oscillations together with the model fit (black line).*

Figure S12a compares the phonon induced dynamics of the $WSe_2$ Peak I and Peak III excitons. Qualitatively both traces exhibit similar behavior, although the Peak III response has a lower signal-to-noise ratio. For completeness, we fit the Peak III dynamics using the driven-oscillator model (Fig. S12b), and present the corresponding Fourier amplitude spectrum and phase in Fig. S12c and d.

In the fitting procedure, the resonance frequencies and linewidths were fixed to the values obtained for Device 1 in Table S1, while only the oscillator strengths were allowed to vary. This constrained fit reproduces the measured dynamics. From the extracted oscillator strengths, we find that $f_1$ contributes approximately 61% of the total oscillator strength, while $f_2$ and $f_3$ contribute 10% and 29%, respectively.

**First-principles atomistic and electronic structure simulations:**

To obtain the equilibrium structure and phonon spectrum of the angle-aligned $WSe_2/WS_2$ heterobilayer, we first constructed a rigid bilayer supercell composed of 25 × 25 $WSe_2$ and 26 × 26 $WS_2$. The atomic structure was then relaxed using LAMMPS[1] with Stillinger–Weber potentials for intralayer interactions and the Kolmogorov–Crespi potential for interlayer interactions[2–4], until the residual forces fell below 1×10-10 eV/ Å. Phonon calculations were subsequently performed on the relaxed structure using the finite-displacement method implemented in phonopy [5], with a maximum displacement amplitude of 0.015 Å and Γ-point sampling. From the resulting force-constant matrix, we computed the phonon density of states on a 12 × 12 q-grid and the phonon dispersion along high-symmetry lines of the moiré Brillouin zone.

Top layer: $WSe_2$
Bottom layer: $WS_2$

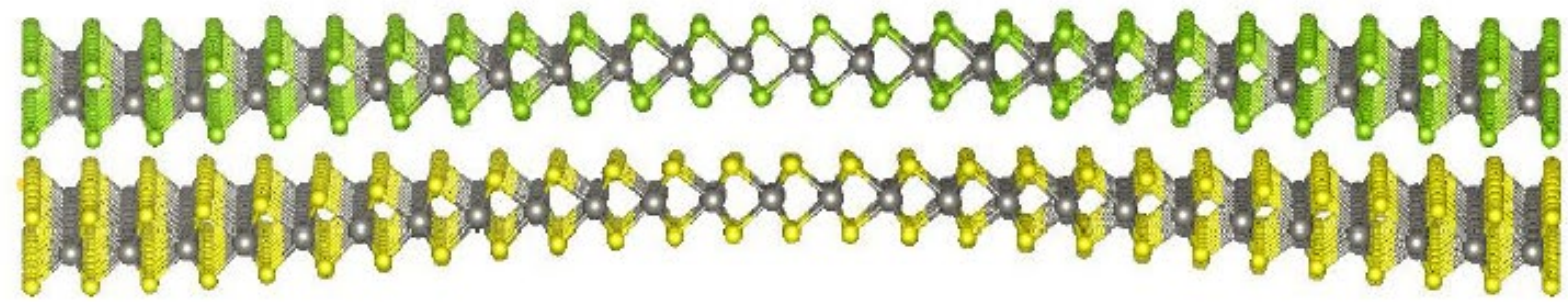

***Figure S13*** *Buckled 3D structure of the angle-aligned $WS_2/WSe_2$ heterobilayer moire superlattice.*

To analyze the effect of the phonon distortions on the electronic structure, we applied finite-difference phonon displacements based on the calculated eigenvectors and carried out DFT calculations using SIESTA[6] with the generalized gradient approximation (GGA) for the exchange–correlation functional[7] and optimized norm-conserving Vanderbilt pseudopotentials from the PseudoDojo library[8,9]. To better describe the wavefunction tails in the slab geometry, one diffuse orbital was added to the double-zeta polarized basis for Se and S atoms. All self-consistent calculations employed Γ -point sampling and a real-space mesh cutoff of 160 Ry.

In the frozen-phonon phonon-displacement calculations, we applied the phonon perturbations using the same phonon normal-coordinate amplitudes, $Q = 1.0 \times 10^3$ and $2.0\ \times 10^3\ a_0\sqrt{m_e}$, and extracted the corresponding changes in the Kohn–Sham eigenvalues. Furthermore, to isolate the electronic states associated with each individual layer more clearly, the frozen-phonon calculations were carried out using isolated monolayer geometries. This treatment is justified by the negligible interlayer hybridization of the relevant band-edge states at the $K$ valleys.

**Breathing-Mode Projection Analysis of Moiré Phonons:**

In this work, our measurement is driven by an out-of-plane coherent wavepacket launched from the few layer graphite transducer and is therefore expected to couple most efficiently to the phonon modes of the sample with significant amplitude along the coordinate of the collective interlayer breathing motion. To isolate such modes, we project the calculated phonon eigenvectors of the bilayer system onto the zero-in-plane-momentum interlayer breathing-mode pattern then analyze the squared breathing-mode weight $P_B^2$. This analysis is similar to the approach used in Ref. [10], where moiré phonon modes with finite layer-breathing or layer-shear character were associated with Raman-active responses. For a phonon mode with branch index $\nu$ and crystal momentum $\mathbf{q}$, we define the breathing-mode projection as

$$P_B = \frac{\sum_{\kappa\alpha} p_{\kappa\alpha}\, \eta_{\kappa\alpha,\nu}(\mathbf{q})}{\sqrt{\sum_{\kappa\alpha} |p_{\kappa\alpha}|^2}},$$

where $\eta_{\kappa\alpha,\nu}(\mathbf{q})$ is the normalized phonon polarization vector, $\kappa$ labels the atoms within the unit cell, and $\alpha$ labels the Cartesian coordinate directions. The breathing-mode projection vector $p_{\kappa\alpha}$ is chosen to model an idealized interlayer breathing motion,

$$p_{\kappa\alpha} = \begin{cases} \sqrt{m_\kappa}, & \alpha = z, \text{and atom } \kappa \text{ belongs to the top layer,} \\ -\sqrt{m_\kappa}, & \alpha = z, \text{and atom } \kappa \text{ belongs to the bottom layer,} \\ 0, & \text{otherwise,} \end{cases}$$

where $m_\kappa$ is the mass of the atom $\kappa$. This choice corresponds to a rigid out-of-plane displacement in which the two layers move in opposite directions and therefore provides a simple measure of the overlap of a given phonon eigenmode with the interlayer breathing coordinate driven by the incident phonon wavepacket.

The four $\mathbf{q}$=0 leading modes in the experimentally relevant frequency window occur at 0.62, 0.65, 0.82, and 0.95 THz, and these are the modes analyzed in the main text. Table S3 lists the leading zone-center modes below 1.5 THz ranked by $P_B^2$, together with representative measures of their in-plane and out-of-plane displacement amplitudes. Here, we additionally analyze phonon mode $\nu$ = 131, which has a frequency of 0.96 THz. This mode is nearly degenerate with mode D at 0.95 THz and exhibits similar quantitative characteristics. From the atomic displacement pattern shown in Fig. S14, we identify it as a rotational pair of mode D, in the sense that its displacement pattern is approximately rotated by 60° around AA stacking regions relative to that of mode D, resulting in a different localization within the moiré unit cell. We further find that its coupling strength is comparable to that of mode D, indicating that it may contribute similarly to the resonances. The remaining modes have smaller breathing-mode weight and generally weaker in-plane reconstruction and are therefore not expected to contribute significantly to the measured excitonic response.

| $\nu$ | Frequency (THz) | $P_B^2$ | $\max\lvert\eta_{xy}^{WSe_2}\rvert$ | $\max\lvert\eta_{xy}^{WS_2}\rvert$ | $\max\lvert\eta_{z}^{WSe_2}\rvert$ | $\max\lvert\eta_{z}^{WS_2}\rvert$ |
|---|---|---|---|---|---|---|
| **81 (C)** | **0.82** | 0.282 | 0.007 | 0.010 | 0.056 | 0.060 |
| **53 (B)** | **0.65** | 0.175 | 0.013 | 0.039 | 0.023 | 0.025 |
| **47 (A)** | **0.62** | 0.113 | 0.038 | 0.008 | 0.024 | 0.024 |
| **128 (D)** | **0.95** | 0.106 | 0.024 | 0.037 | 0.039 | 0.063 |
| 77 | 0.78 | 0.094 | 0.010 | 0.011 | 0.038 | 0.090 |
| **131** | **0.96** | 0.083 | 0.023 | 0.037 | 0.038 | 0.065 |
| 125 | 0.94 | 0.046 | 0.016 | 0.016 | 0.053 | 0.095 |
| 101 | 0.87 | 0.027 | 0.011 | 0.006 | 0.066 | 0.074 |
| 82 | 0.82 | 0.018 | 0.007 | 0.011 | 0.062 | 0.065 |

***Table S3*** *Quantitative summary of selected zone-center moiré phonon modes in aligned $WSe_2/WS_2$. Listed are the phonon frequency, squared breathing-mode projection $P_B^2$ and representative measures of the in-plane and out-of-plane displacement amplitudes in the $WSe_2$ and $WS_2$ layers. The modes are ordered by decreasing $P_B^2$.*

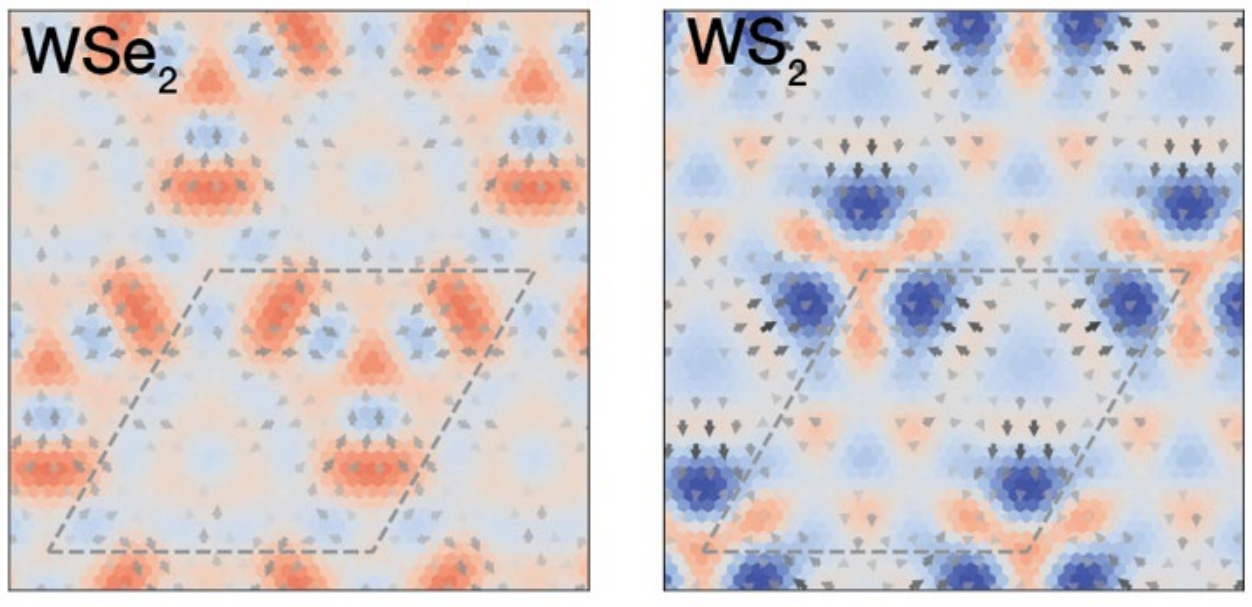

***Figure S14*** *Atomic displacement pattern of phonon mode $\nu$ = 131. Arrows indicate the in-plane atomic displacements, while the blue-to-red color map shows the out-of-plane displacement component, with red (blue) corresponding to motion along the +z (-z) direction.*

After identifying the relevant phonon modes ranked by $P_B^2$, we fix the sign of each phonon eigenvector to compare the signs of the phonon-induced electronic responses across phonon modes. Here we adopt a sign convention guided by the device geometry and excitation scheme. The coherent acoustic wavepacket is injected from the bottom side of the heterostructure and therefore couples first to out-of-plane motion near the lowest point of the relaxed moiré structure, located around the AA stacking region of the $WS_2$ layer adjacent to the bottom hBN. For each zone-center phonon mode, we choose the eigenvector sign such that the atoms in this region move in the same out-of-plane direction. Under this convention, the signs of the frozen-phonon band-gap shifts can be compared consistently across modes and related to the experimentally inferred signs of the excitonic response.

**Electron-phonon coupling analysis on $WSe_2$ electronic states:**

We present the diagonal electron-phonon coupling matrix elements $g_{\mathbf{nk}}^{(\nu)} = \sqrt{\frac{\hbar}{2\omega_\nu}}\frac{\partial\varepsilon_{\mathbf{nk}}}{\partial Q_\nu}$ in Table S4 for the $WSe_2$ states relevant to Peak I. Across these low-energy $WSe_2$ band states, mode C consistently exhibits the smallest matrix elements, while modes A and D show substantially stronger coupling, consistent with the trends inferred from the frozen-phonon gap shifts (Fig 4c).

| Band | **A** 0.62 THz | **B** 0.65 THz | **C** 0.82 THz | **D** 0.95 THz |
|---|---|---|---|---|
| $c_1$ | -0.774 | 0.175 | -0.056 | -0.284 |
| $v_1$ | 0.535 | -0.204 | -0.050 | 0.175 |
| $v_2$ | 0.133 | -0.040 | -0.008 | 0.004 |

***Table S4*** *Diagonal part of the electron phonon matrix elements* $g_{nk}^{(\nu)}$ *in meV at electron crystal momentum* $\boldsymbol{k} = \gamma$ *for the low-energy* $WSe_2$ *band states and selected zone-center moiré phonon modes.*

**Electron-phonon coupling analysis on $WS_2$ electronic states:**

For completeness, we also computed the diagonal electron-phonon matrix elements for several low-energy $WS_2$ states. Although the dominant band-to-band transitions underlying the measured $WS_2$ excitonic response are less clearly established, these results show the same qualitative trend, namely that phonon modes with stronger in-plane displacements in the $WS_2$ layer exhibit stronger coupling to the relevant low-energy electronic states.

| Band | **A**-0.62 THz | **B**-0.65 THz | **C**-0.82 THz | **D**-0.95 THz |
|---|---|---|---|---|
| $c_1$ | 0.280 | -1.407 | -0.016 | -0.488 |
| $v_1$ | -0.151 | 0.785 | -0.095 | 0.460 |
| $v_2$ | -0.041 | 0.278 | -0.054 | -0.208 |

***Table S5*** *Diagonal part of the electron phonon matrix elements* $g_{nk}^{(\nu)}$ *in meV at electron crystal momentum* $\boldsymbol{k} = \gamma$ *for the low-energy* $WS_2$ *band states and selected zone-center moiré phonon modes.*

**Phonon-induced local strain fields:**

To further clarify the microscopic origin of the mode-dependent exciton-phonon coupling, in Fig. S15 we plot the phonon-induced change in the local strain field for the four zone-center moiré phonon modes discussed in the main text. The strain maps provide a direct visualization of how strongly each phonon perturbs the local moiré reconstruction within each layer. Consistent with the displacement patterns shown in Fig. 4b, Modes A (0.62 THz) and B (0.65 THz) produce pronounced strain modulation predominantly in the $WSe_2$ and $WS_2$ layers, respectively, reflecting their layer-selective character. Mode D (0.95 THz) also induces substantial strain variation, distributed across both layers. By contrast, Mode C (0.82 THz) generates only a very weak strain-field change in either layer. This result supports the interpretation that the 0.82 THz mode has the weakest coupling to the low-energy electronic structure: although it retains finite breathing-mode projection and can in principle be driven by the incident phonon wavepacket, it produces only minimal in-plane lattice reconstruction and therefore only weakly modulates the moiré potential.

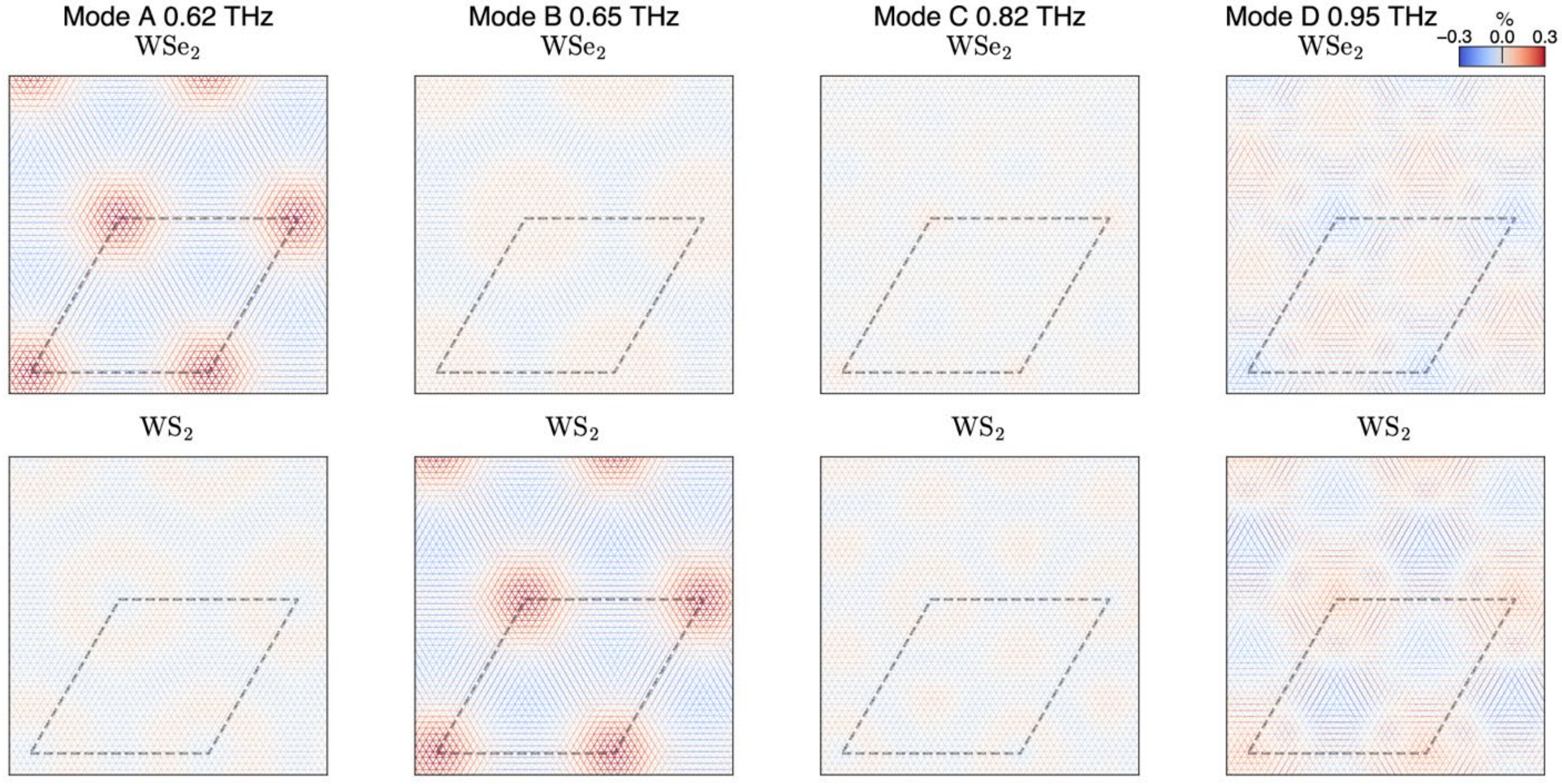

***Figure S15*** *Spatial maps of the change in local strain induced by the four selected zone-center moiré phonon modes: Mode A (0.62 THz), Mode B (0.65 THz), Mode C (0.82 THz), and Mode D (0.95 THz). The top and bottom rows show the strain response in the* $WSe_2$ *and* $WS_2$ *layers, respectively. The dashed parallelogram marks the moiré unit cell. Red and blue colors indicate positive and negative local strain variations. For all modes, the distorted structures were generated using the same phonon normal-coordinate amplitude,* $Q = 1.0 \times 10^3\ a_0\sqrt{m_e}$*, so the relative magnitude of the strain fields directly reflects differences in the phonon eigenvectors. Modes A and B generate pronounced strain modulation predominantly in one layer, consistent with their layer-selective character. Mode D produces appreciable strain variation in both layers. In contrast, Mode C induces the weakest strain-field modulation in both layers, supporting the conclusion that this mode couples only weakly to the low-energy moiré electronic structure despite its finite breathing-mode character.*

**Electronic band structures of isolated moiré $WS_2$ and $WSe_2$:**

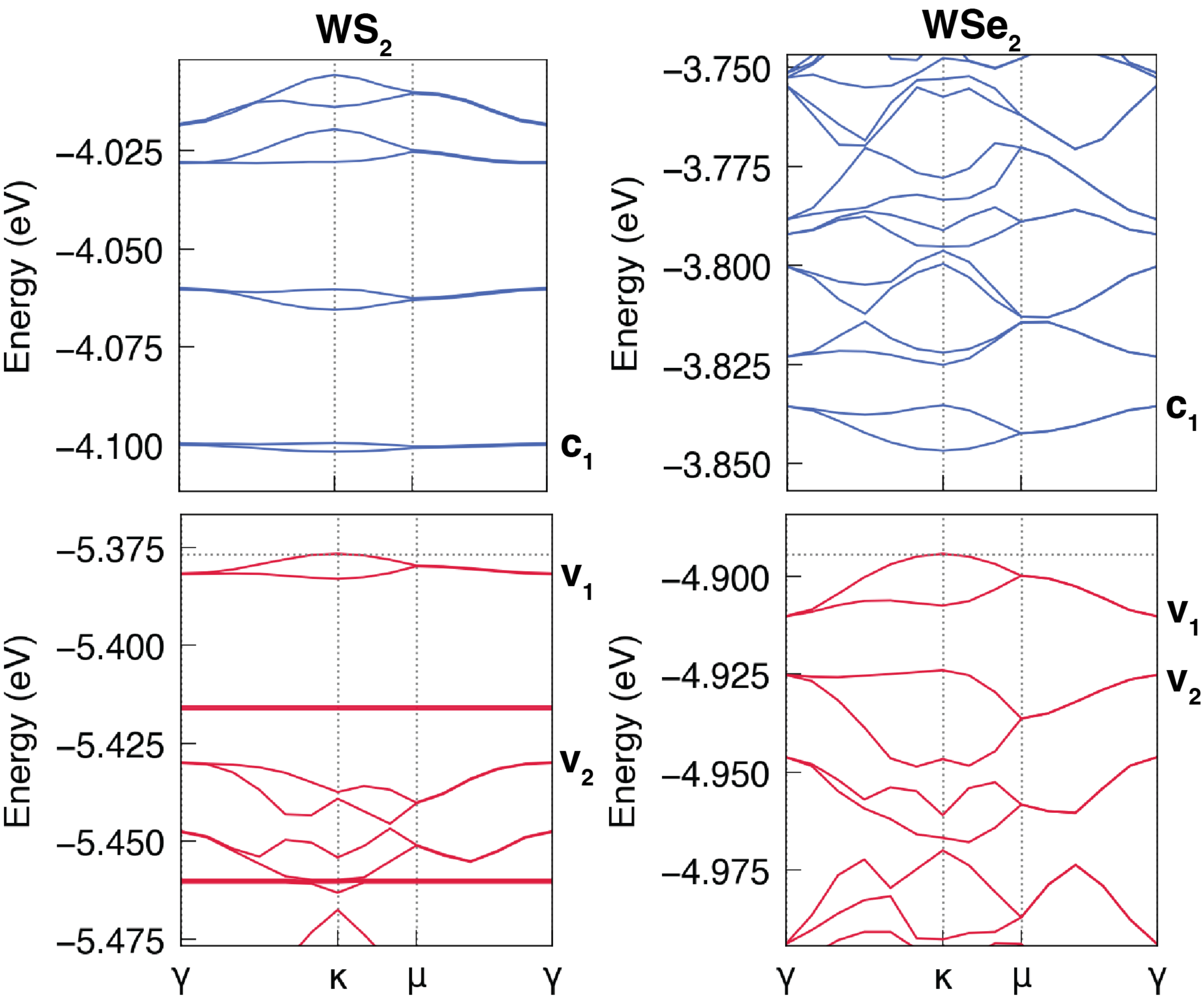

***Figure S16*** *Electronic band structures of isolated moiré $WS_2$ (left panels) and $WSe_2$ (right panels) from DFT calculations. The top panels show the bands near the conduction-band edge, and the bottom panels show those near the valence-band edge. The zero of the energy scale is set at the vacuum level. The states used in the frozen-phonon analysis are labeled. For the $WS_2$ valence bands, we exclude the flat bands derived from $\Gamma$-valley states and consider only the K-valley states that are optically active.*

## Supplementary References